\documentclass[apj,numberedappendix]{emulateapj}
\journalinfo{Accepted to the Astrophysical Journal}

\usepackage{rotating}

\shorttitle{The Mass-Size Relation from Clouds to Cores. II. Solar
    Neighborhood Clouds}
\shortauthors{Kauffmann et al.}

\begin{document}

\title{The Mass-Size Relation from Clouds to Cores. II. Solar
  Neighborhood Clouds}

\author{J.\ Kauffmann\altaffilmark{1,2,$\star$}, T.\
  Pillai\altaffilmark{2,$\star$}, R.\
  Shetty\altaffilmark{1,2,$\star$}, P.\ C.\ Myers\altaffilmark{2}, \&
  A.\ A.\ Goodman\altaffilmark{1,2}}

\altaffiltext{1}{Initiative in Innovative Computing (IIC), 60 Oxford
  Street, Cambridge, MA 02138, USA}
\altaffiltext{2}{Harvard-Smithsonian Center for Astrophysics, 60
  Garden Street, Cambridge, MA 02138, USA}
\altaffiltext{$\star$}{present addresses: Jens Kauffmann, NPP Fellow,
  Jet Propulsion Laboratory, 4800 Oak Grove Drive, Pasadena, CA 91109,
  USA; Thushara Pillai, CARMA Fellow, Caltech Astronomy Department,
  1200 East California Blvd., Pasadena, CA 91125, USA; Rahul Shetty,
  Zentrum f\"ur Astronomie der Universit\"at Heidelberg, Institut
  f\"ur Theoretische Astrophysik, Albert-Ueberle-Str.\ 2, D-69120
  Heidelberg, Germany}

\email{jens.kauffmann@jpl.nasa.gov}

\begin{abstract}
  We measure the mass and size of cloud fragments in several molecular
  clouds continuously over a wide range of spatial scales
  ($0.05 \lesssim r / {\rm pc} \lesssim 3$). Based on the recently developed
  ``dendrogram-technique'', this characterizes dense cores as
  well as the enveloping clouds. ``Larson's 3$^{\rm rd}$ Law'' of
  constant column density, $m(r) \propto r^2$, is not well suited
  to describe the derived mass-size data. Solar neighborhood clouds
  not forming massive stars ($\lesssim 10 \, M_{\sun}$; Pipe Nebula,
  Taurus, Perseus, and Ophiuchus) obey
  $$ m(r) \le 870 \, M_{\sun} \, (r / {\rm pc})^{1.33} \, . $$
  In contrast to this, clouds forming massive stars (Orion~A,
  G10.15$-$0.34,  G11.11$-$0.12) do exceed the aforementioned
  relation. Thus, this limiting mass-size relation may approximate a
  threshold for the formation of massive stars. Across all clouds,
  cluster-forming cloud fragments are found to be---at given
  radius---more massive than fragments devoid of clusters. The
  cluster-bearing fragments are found to roughly obey a mass-size law
  $m \propto r^{1.27}$ (where the exponent is highly uncertain in any
  given cloud, but is certainly smaller than 1.5).
\end{abstract}


\maketitle

\begin{figure*}
\includegraphics[scale=0.55,bb=20 54 361 386,clip]{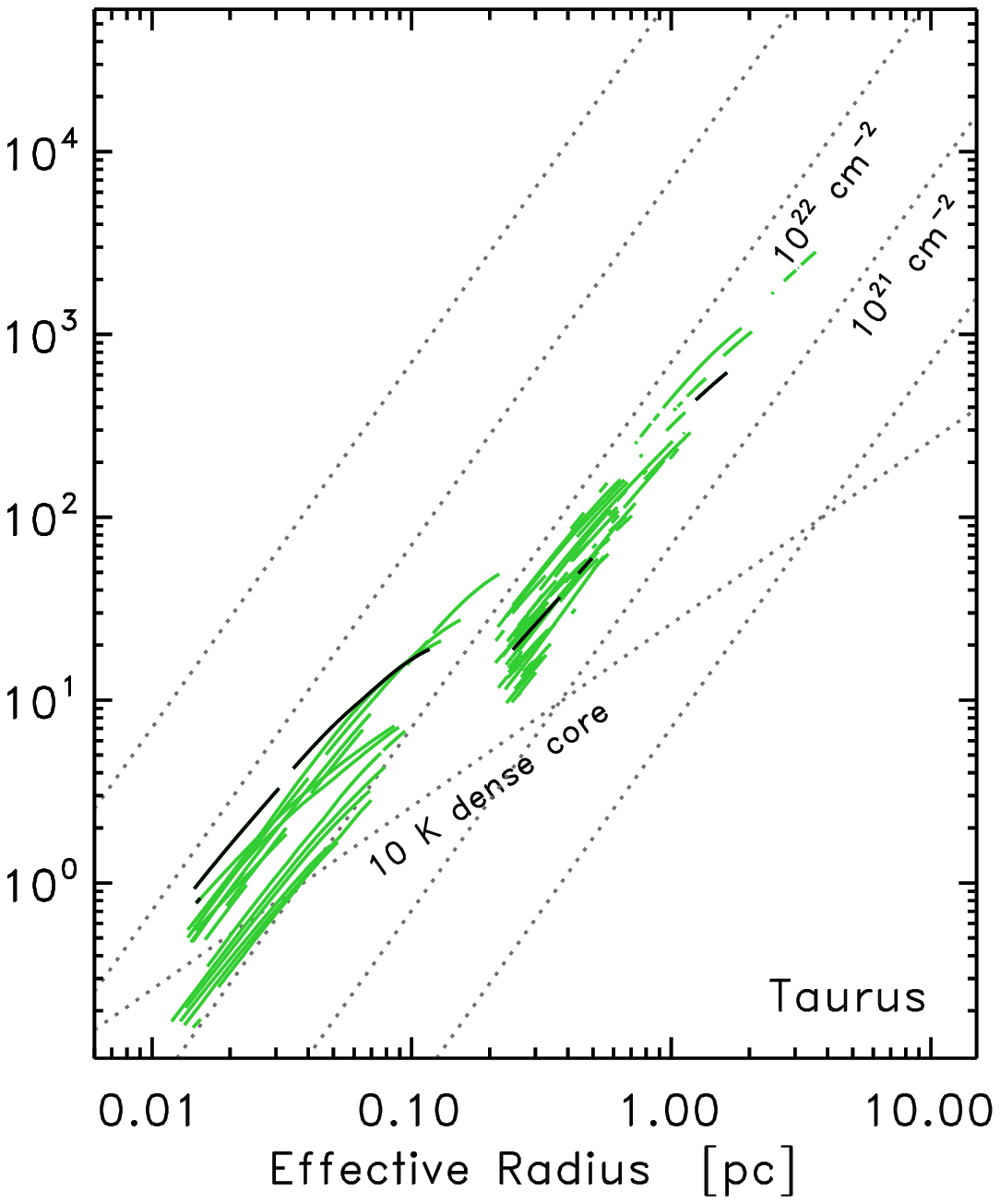}
\includegraphics[scale=0.55,bb=76 54 361 386,clip]{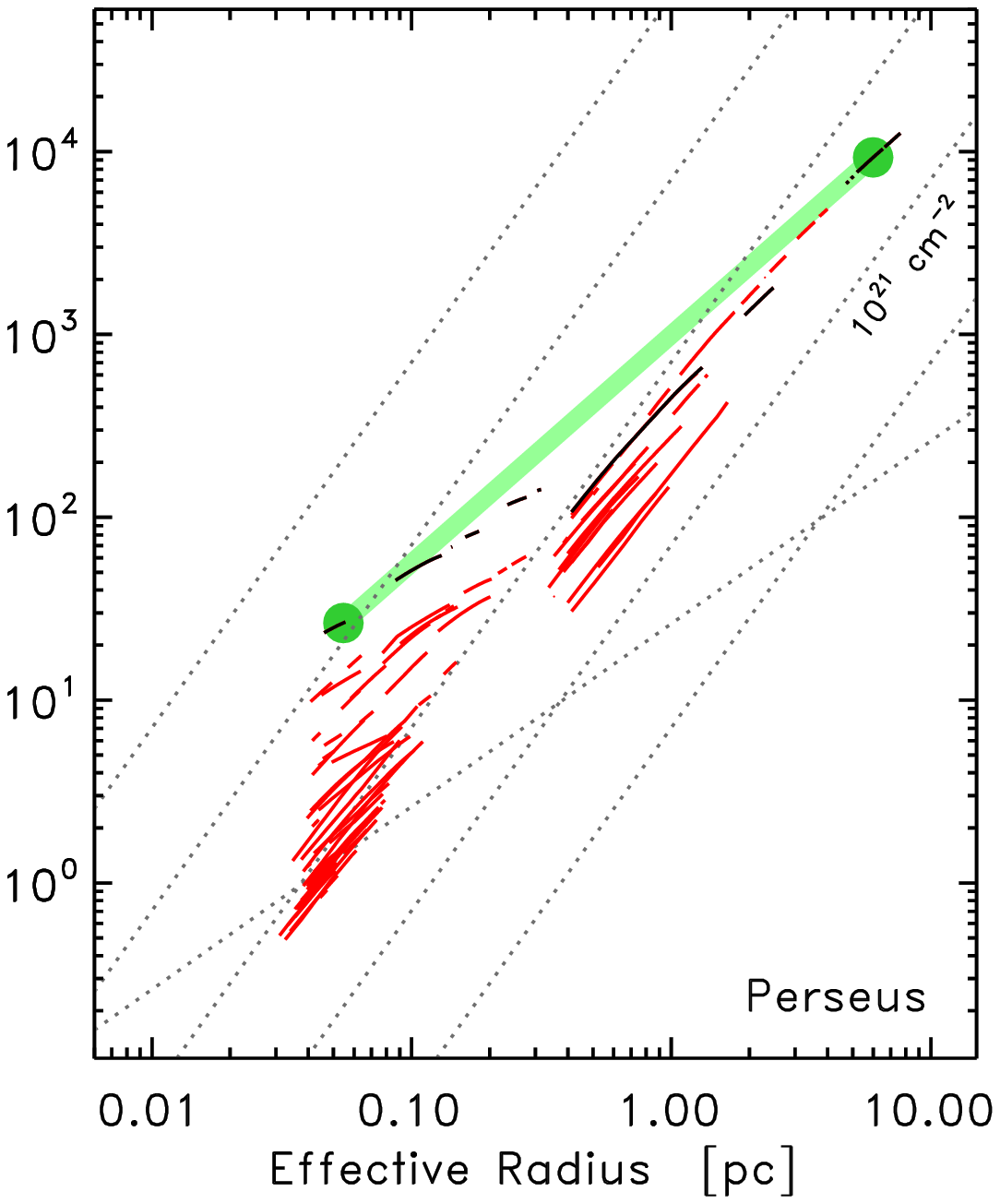}
\includegraphics[scale=0.55,bb=76 54 361 386,clip]{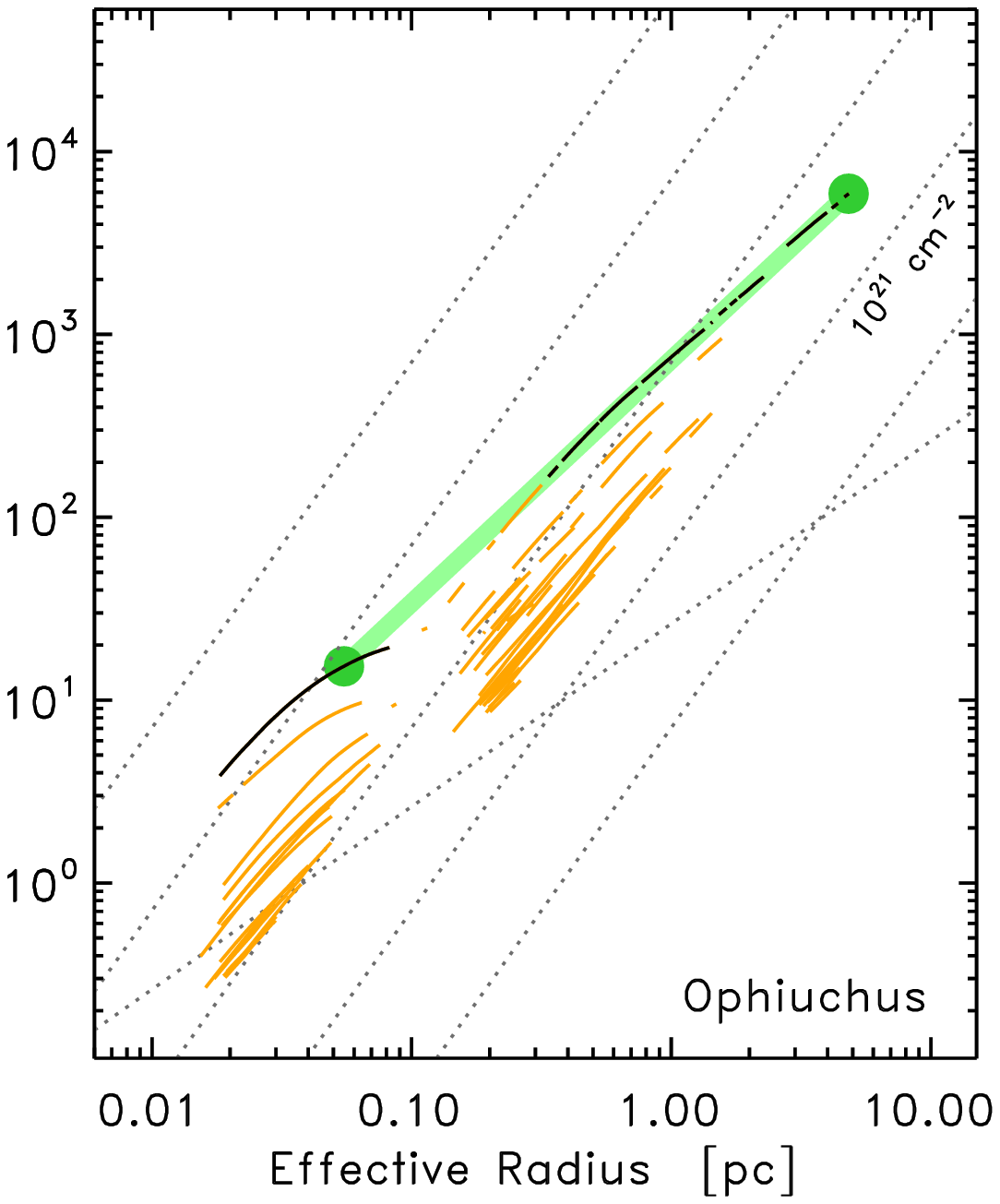}\\
\includegraphics[scale=0.55,bb=20 10 361 386,clip]{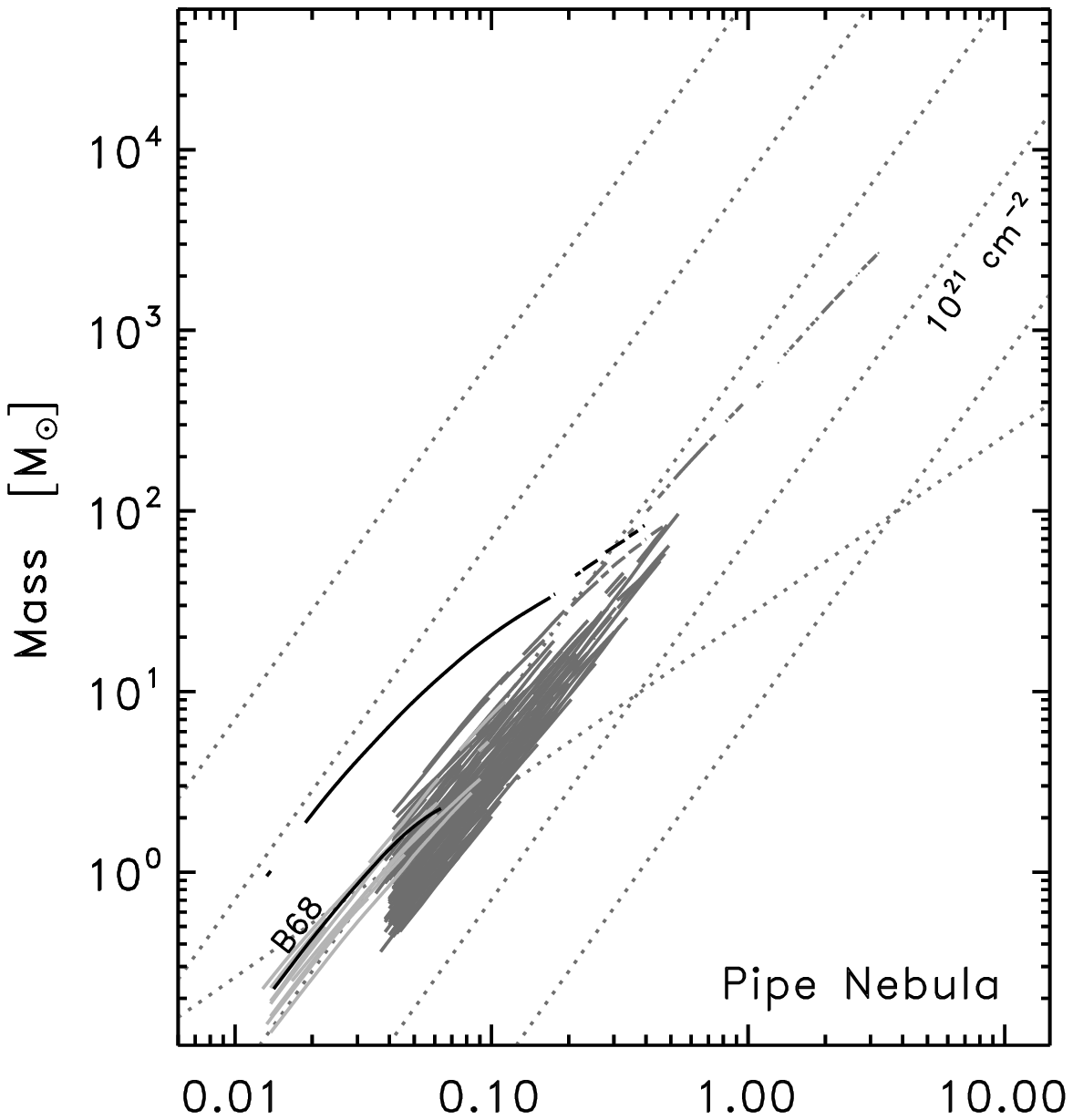}
\includegraphics[scale=0.55,bb=76 10 361 386,clip]{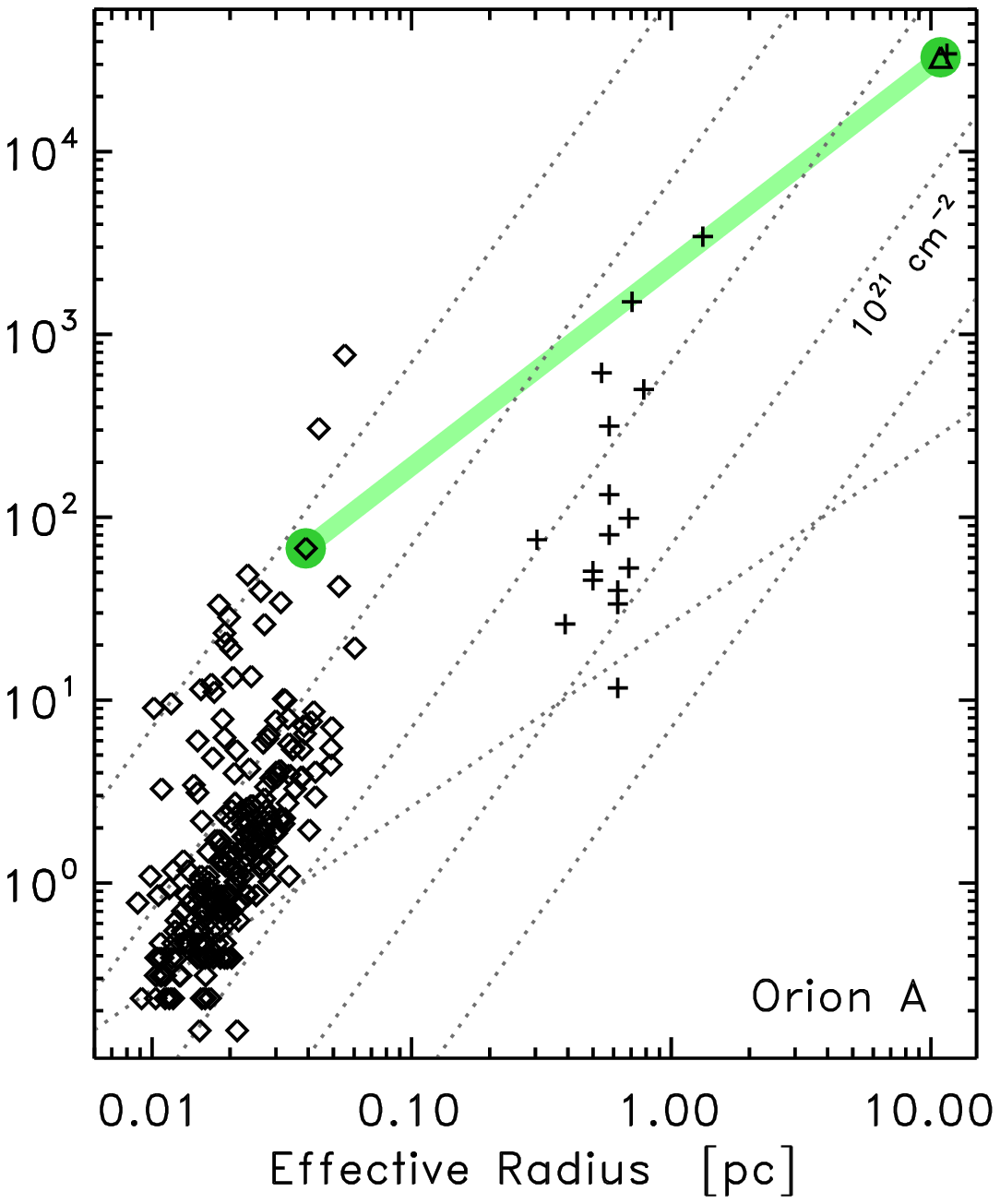}
\includegraphics[scale=0.55,bb=76 10 361 386,clip]{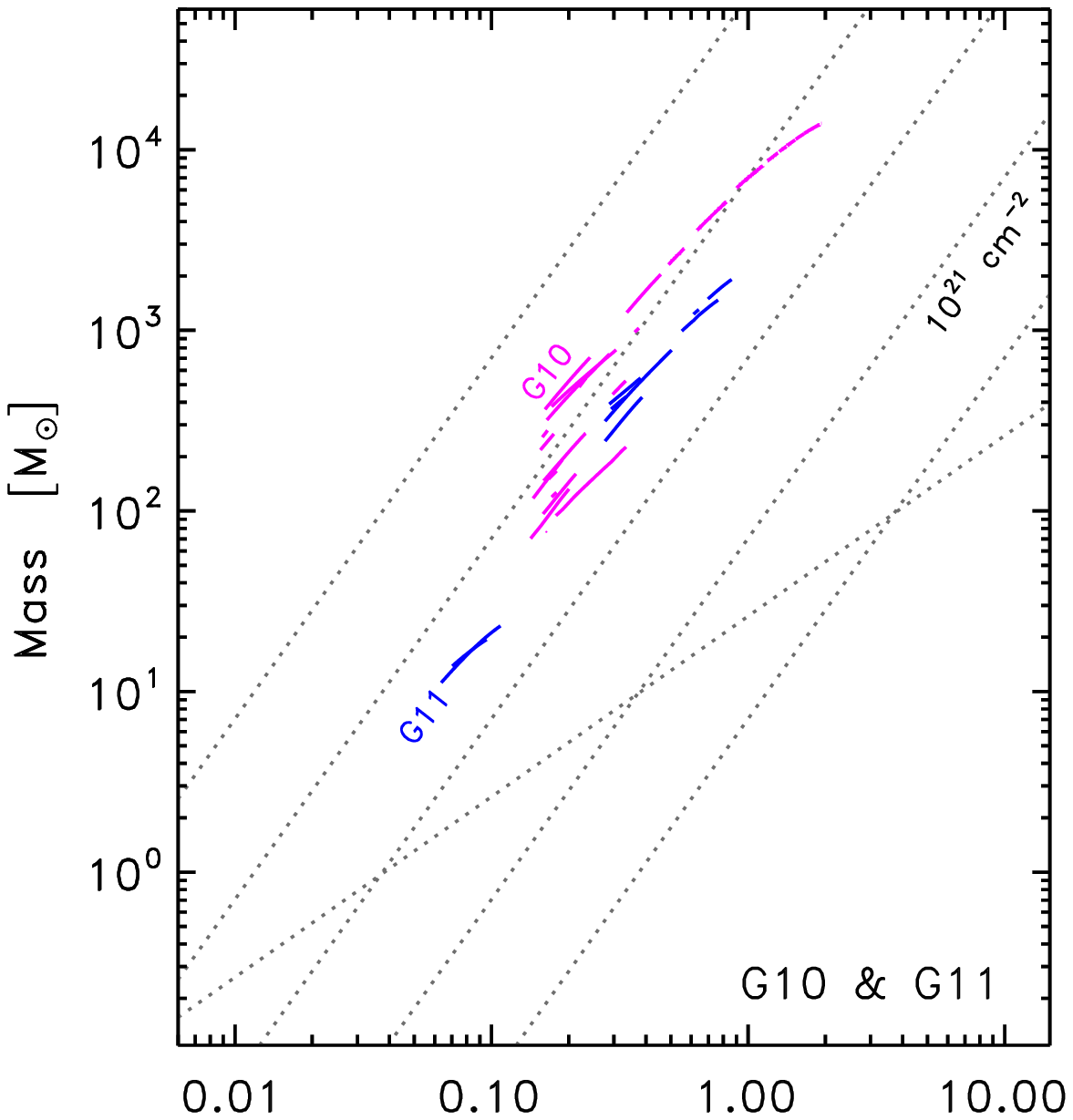}
\caption{Mass-size relations for the clouds listed in Table
  \ref{tab:included-data}. In most clouds, data from two different
  observational techniques (e.g., dust extinction and emission) are
  combined to provide a comprehensive picture probing a wide range of
  spatial scales. \emph{Solid black lines} highlight cloud regions
  containing the most massive cloud fragment found for small
  radii. \emph{Green lines starting in circles} indicate the
  global mass-size relations discussed in Section
  \ref{sec:slope-definitions}. The other \emph{dotted lines} give
  reference mass-size relations as discussed in Sec.\
  \ref{sec:reference}. In Orion~A, published data extracted using
  CLUMPFIND-like approaches are plotted (\emph{diamonds} mark SCUBA
  data, \emph{crosses} indicate CO observations) instead of using the
  contour-based scheme employed here. Further, the \emph{triangle}
  indicates the large-scale Orion~A extinction mass measurement
  discussed in the text.\label{fig:cloud-sample}}
\end{figure*}

\section{Introduction}
Most of our present understanding of star formation processes is based
on detailed studies of solar neighborhood molecular clouds (closer
$\sim 500 ~ \rm pc$). To this end past research has, e.g., studied the
masses and sizes of dense cores in molecular clouds
($\lesssim 0.1 ~ \rm pc$ size) such as Perseus, Taurus, Ophiuchus,
Orion, and the Pipe Nebula (e.g.\ \citealt{motte1998:ophiuchus},
\citealt{johnstone2000:rho_ophiuchi}, \citealt{hatchell2005:perseus},
\citealt{enoch2007:cloud_comparison}). Further research studied clumps
in these clouds (some $0.1 ~ \rm pc$) and the clouds
($\gtrsim 10 ~ \rm pc$) containing the cores (e.g.,
\citealt{williams1994:clumpfind}, \citealt{cambresy1999:extinction},
\citealt{kirk2006:scuba-perseus}; see \citealt{williams2000:pp_iv} for
definitions of cores, clumps, and clouds). This research does,
however, not probe the \emph{relation} between the properties of
cores, clumps, and clouds: traditionally, every domain is
characterized and analyzed separately. As a result, even in the solar
neighborhood, it is still not known how the core densities (and thus
star-formation properties) relate to the state of the surrounding
cloud.

To be precise, we do in principle know a bit about the relation
between cloud structure at large and small scale. For structure within
molecular clouds, \citet{larson1981:linewidth_size} concluded (in his
Eq.\ 5) that the mass contained within the radius $r$ obeys a
power-law,
\begin{equation}
m(r) = 460 \, M_{\sun} \, (r / {\rm pc})^{1.9} \, .
\label{eq:mass-size-larson}
\end{equation}
Most subsequent work refers to this relation as ``Larson's
3$^{\rm rd}$ law'', and replaces the original result with
$m(r) \propto r^2$ (e.g., \citealt{mckee2007:review}). This ``law of
constant column density'' (with respect to scale, $r$) is now
considered one of the fundamental properties of molecular cloud
structure (e.g., \citealt{ballesteros-paredes2007:ppv},
\citealt{mckee2007:review}, \citealt{bergin2007:dense-core-review}).
We do, however, not know whether this relation is still consistent
with up-to-date column density maps of molecular clouds.\medskip

Part I of the present series \citep{kauffmann2010:mass-size-i}
describes a new technique to extract mass-size relations from cloud
maps. It is based on ``dendrograms'', a tree-based segmentation of
cloud structure \citep{rosolowsky2008:dendrograms}. Here, we employ
this technique to study the molecular clouds in Perseus, Taurus,
Ophiuchus, Orion, and the Pipe Nebula. To illustrate the properties of
more massive clouds, we also include data for two more distant clouds
of high density (farther than $2 ~ \rm kpc$; G10.15$-$0.34 and
G11.11$-$0.12).\medskip

The present paper summarizes the analysis method in Sec.\
\ref{sec:method}. The main quantitative analysis of the maps is
presented in Sec.\ \ref{sec:sample-analysis}. Section
\ref{sec:interpretation} systematizes the results and interprets them
in the context of our present knowledge of star formation
regions. This discussion is supported by model calculations in
Appendices \ref{sec-app:mass-size-models} and
\ref{sec-app:polytropes}. We conclude with a summary in Sec.\
\ref{sec:summary}.

\section{Method \& Data}
\label{sec:method}

\subsection{Data Processing}
\subsubsection{Basic Map Analysis}
Our basic analysis approach is summarized in Sec.\ 2.1 of part I, and
illustrated in Fig.\ 1 of the same paper. In essence, starting from a
set of local maxima, we contour a given column density map at all
levels possible. For every contour, we derive the enclosed mass and
area, $A$. Following the terminology of
\citet{peretto2009:irdc-catalogue}, we define ``cloud fragments'' in
the maps as such regions enclosed by a continuous column density
contour. The area is used to derive effective radii,
\begin{equation}
r = (A / \pi)^{1/2} \, .
\end{equation}
Subsequent contours are usually nested in the map. This defines a
relation between measurements. This essentially yields series of
mass-size measurements. In the plots shown in this paper, such series
are drawn using continuous lines (e.g., Fig.\ \ref{fig:cloud-sample}).

In practice, the processing is implemented using the dendrogram map
analysis technique introduced by
\citet{rosolowsky2008:dendrograms}. As a bonus, this also yields
diagnostic diagrams on the cloud hierarchy. We do not use this feature
in the present paper, though.

In our analysis, we reject all cloud fragments that have a diameter
(i.e., $2 r$) smaller than the map resolution. Further, we require a
minimum contrast between the local maxima used to seed the
contouring. Here, this limit is set to the noise level times a factor
3. Further, we do not characterize column densities below a certain
threshold. As discussed below, Table \ref{tab:included-data} lists
these parameters for each map studied here.

\subsubsection{Calculation of Mass-Size
  Slopes\label{sec:slope-definitions}}
Part I demonstrates that mass-size data can conveniently be
characterized using power-laws of the form
\begin{equation}
m(r) = m_0 \, (r / {\rm pc})^b \, .
\label{eq:power-law}
\end{equation}
These are characterized by slope, $b$, and intercept, $m_0$. As shown
in part I, slopes can be derived using various methods. As
demonstrated in part I, $b \le 2$ in our work, since $b > 2$ would
imply an increase of the mean column density with radius.

We define global slopes to capture trends between small and large
spatial scales (Sec.\ 4.3 in part I). To calculate
these, we derive the maximum fragment masses, $m_{\rm max}$, observed
at radii of $r_{\rm sm} = 0.05 ~ \rm pc$ and
$r_{\rm lg} = 5.0 ~ \rm pc$. (These radii are chosen to permit
comparison between clouds, as becomes more obvious below.) Based on
these, we derive the global slope, 
\begin{equation}
b_{\rm glob} = \frac{
  \ln[m_{\rm max}(r_{\rm lg}) / m_{\rm max}(r_{\rm sm})]
}{
  \ln[r_{\rm lg} / r_{\rm sm}]
} \, .
\label{eq:slope-global}
\end{equation}
As illustrated in Fig.\ \ref{fig:cloud-sample}, this slope is
defined such that $m(r) \propto r^{b_{\rm glob}}$ connects the mass
and size measurements at $0.05 ~ \rm pc$ and $5.0 ~ \rm pc$ radius
(for appropriate intercept).

To characterize trends at a given spatial scale, we use tangential
slopes (Sec.\ 4.4 in part I). These are derived infinitesimally at a
given radius,
\begin{equation}
b(r) =
\left.
\frac{{\rm d} \, \ln(m[r'])}{{\rm d} \, \ln(r')}
\right|_{r' = r} \, .
\label{eq:slopes_tangential}
\end{equation}
As illustrated in Fig.\ 7 of part I, the measured slopes are smoothed
and filtered to improve the data quality.

\subsubsection{Reference Mass-Size Relations\label{sec:reference}}
Section 3.1 of part I introduces various reference mass-size
relations. These can be used to navigate more intuitively within the
data space. They are thus featured in most figures of the present
paper (e.g., Fig.\ \ref{fig:cloud-sample}). Specifically, lines of
constant mean column density obey a mass-size law
$m(r) \propto r^2$. Our plots contain relations for mean column
densities separated by factors of 10. Static equilibrium models of
isothermal spheres, on the other hand, obey $m(r) \propto r$. We draw
a model relation for a gas temperature of $10 ~ \rm K$.

We stress that we obtain two-dimensional mass-size relations from
column density maps. These are related to, but not identical with,
mass-size laws obtained from three-dimensional density maps. This is
further explored in Sec.\ \ref{sec:density-pressure}.

\subsection{Combining Data for several Clouds}
Part I describes in detail how dust extinction and emission maps can
be used to derive column density maps for molecular cloud
complexes. Once calibrated to a common mass conversion scale (Sec.\
4.2 of part I), dust extinction and emission data for a given cloud
can be combined to probe the masses and sizes across a vast range of
spatial scales. In part I, we did this for Perseus only. Here, we
combine our Perseus data with maps of other clouds. We do this by
repeating the analysis already carried our for Perseus.

We start with a survey of a sample of well-studied nearby clouds that
are essentially devoid of high-mass stars. Besides the Perseus
molecular cloud, those in Taurus, Ophiuchus, and the Pipe Nebula are
examined here. As a general reference to clouds also forming high-mass
stars, we include the Orion~A cloud. Then, the more remote
G10.15$-$0.34 (hereafter G10; $\sim 2.1 ~ \rm kpc$) and G11.11$-$0.12
complexes (hereafter G11; $\sim 3.6 ~ \rm kpc$) are studied to build
a first bridge towards the study of relatively distant sites of high
mass star formation (see \citet{pillai2007:initial-conditions} for
distances and references; G10 is further discussed by
\citet{wood1989:uchii} and \citealt{thompson2006:scuba-hii}, while
\citet{pillai2006:g11} study G11). Note that G11 is an Infrared Dark
Cloud (IRDC; see \citealt{menten2005:irdc-review} and
\citealt{beuther2007:massive-sf} for reviews).

A data summary for our sample is provided in Table
\ref{tab:included-data}. This includes parameters used for the source
extraction.

\begin{table*}
\caption{Data included in this study.\label{tab:included-data}}
\begin{center}
\begin{tabular}{llllllllllll}
\hline\hline
\rule[-1.0ex]{0ex}{4.5ex}Region / Distance & Data & Resolution &
Noise & Rejection Threshold & Reference\\
\rule[-1.5ex]{0ex}{4.0ex} & & & mag & mag\\
\hline
\multicolumn{6}{l}{\textit{\rule[0ex]{0ex}{3ex}clouds only forming low-mass stars:}}\\
\rule[0ex]{0ex}{3ex}Taurus & extinction & $\le 6 \arcmin$ & $0.4 $ &
 $2.0 $ & \citet{rowles2009:extinction-maps}\\
$d = 140 ~ \rm pc$ & dust emission (10.0 K) & $20 \arcsec$ & $1.4 $ &
 $4.3 $ & \citet{kauffmann2008:mambo-spitzer}\\
\rule[0ex]{0ex}{3ex}Perseus & extinction & $5 \arcmin$ & $0.4 $ &
 $2.0 $ & \citet{ridge2006:complete-phase_i}\\
$d = 260 ~ \rm pc$ & dust emission (12.5 K) & $30 \arcsec$ & $1.4 $ &
 $4.3 $ & \citet{enoch2006:perseus}\\
\rule[0ex]{0ex}{3ex}Ophiuchus & extinction & $5 \arcmin$ & $0.6 $ &
 $2.0 $ & \citet{ridge2006:complete-phase_i}\\
$d = 120 ~ \rm pc$ & dust emission (12.5 K) & $30 \arcsec$ & $2.5 $ &
 $7.6 $ & \citet{enoch2007:cloud_comparison}\\
\rule[0ex]{0ex}{3ex}Pipe Nebula & extinction & $60 \arcsec$ & $0.5 $ &
 $4.0 $ & \citet{lombardi2006:pipe}\\
$d = 130 ~ \rm pc$\\
\rule[0ex]{0ex}{3ex}B59 & extinction & $20 \arcsec$ & $1.2$ &
 $3.6$ & \citet{roman-zuniga2009:b59-extinction}\\
$d = 130 ~ \rm pc$\\
\rule[0ex]{0ex}{3ex}B68 & dust emission (10.0 K) & $15 \arcsec$ & $1.1$ &
 $3.3$ & J.\ Alves, priv.\ comm.\\
$d = 130 ~ \rm pc$\\
\multicolumn{6}{l}{\textit{\rule[0ex]{0ex}{4.5ex}clouds also forming high-mass stars:}}\\
\rule[0ex]{0ex}{3ex}Orion~A & $\rm ^{13}CO$ & $1 \farcm 7$ & --- &
 --- & \citet{bally1987:orion}\\
$d = 414 ~ \rm pc$ & dust emission (20.0 K) & $14 \arcsec$ & --- &
 --- & \citet{nutter2007:orion}\\
\rule[0ex]{0ex}{3ex}G10.15$-$0.34 & dust emission (20.0 K) & $14 \arcsec$ & $6.4$ &
 $19.2 $ & Thompson et al., in prep.\\
\rule[-1.5ex]{0ex}{0ex}$d = 2100 ~ \rm pc$\\
\rule[0ex]{0ex}{3ex}G11.11$-$0.12 & dust emission (15.0 K) & $14 \arcsec$ & $6.6$ &
 $19.7$ & \citet{carey2000:irdc-submillimeter}\\
\rule[-1.5ex]{0ex}{0ex}$d = 3600 ~ \rm pc$ & dust emission (15.0 K) &
 $4 \arcsec$ & $4.9$ & $14.6$ & Pillai et al., in prep.\\
\hline
\end{tabular}
\end{center}
\end{table*}

\subsubsection{Data and Analysis\label{sec:data-analysis}}
The extinction map analysis for Ophiuchus
\citep{ridge2006:complete-phase_i} and the Pipe Nebula
\citep{lombardi2006:pipe} is analogous to the one for the Perseus
cloud carried out in part I. The visual extinction in the Pipe Nebula
map is reduced by $1.34 ~ {\rm mag}$ (reduction of $0.15 ~ \rm mag$ in
$A_K$), following the analysis by
\citet{lombardi2008:ophiuchus-lupus}.  For Taurus, we use the map by
\citet{rowles2009:extinction-maps}. Its resolution varies throughout
the map. This is analogous to a region-dependent smoothing, which may
affect mass estimates (Sec.\ 4.1 in part I). We thus use this map with
caution. In Taurus \citep{kauffmann2008:mambo-spitzer} and Ophiuchus
\citep{enoch2007:cloud_comparison} the processing of the dust emission
maps follows the Perseus Bolocam analysis of part I. \emph{Note that
  the Taurus MAMBO maps do not cover all of the cloud and may
  therefore give a biased view of the dense core conditions.} Pipe
Nebula regions with high column density must be probed in detail. To
do this, we include a SCUBA map for B68 \citep{alves2001:b68}, which
is kindly provided by J.\
Alves. \citet{roman-zuniga2009:b59-extinction} mapped the Pipe's B59
region in extinction. We process these maps using our analysis
scheme. For all sources, Table \ref{tab:included-data} lists the free
parameters used in our source extraction algorithm.

The resulting mass-size relations form part of Fig.\
\ref{fig:cloud-sample}. To illustrate some aspects of the spatial
cloud structure, Fig.\ \ref{fig:cloud-sample} highlights (in black)
for every cloud the mass-size evolution of the most massive fragment
found at small radii. Note that, in Taurus and the Pipe Nebula, these
fragments do \emph{not} form part of the largest cloud fragments found
in the maps.

In Orion~A, no data suited for a reliable dendrogram structure
analysis of column densities is yet available. Instead, we use the
\citet{rowles2009:extinction-maps} extinction maps to derive a single
mass and size measurement on the largest scales probed by that map. On
smaller scales, the extinction map is not reliable because of too few
background sources. For reference, we also plot mass and size
measurements for $\rm ^{13}CO$ clumps in Orion, as published elsewhere
(\citealt{bally1987:orion}; their Table 1, plus text
statements). These data are, however, not used in the quantitative
analysis. At smaller scales, we fold in published data from SCUBA
studies employing CLUMPFIND-like data analysis methods
\citep{nutter2007:orion}. As illustrated in Fig.\ 4 of part I, at
given radius, these set a lower limit to the maximum
dendrogram-derived masses usually employed here.

For G10 (the data are kindly provided---in advance of publication---by
M.\ Thompson, J.\ Hatchell, and F.\ Wyrowski) and G11
\citep{carey2000:irdc-submillimeter}, we run our contour analysis on
SCUBA maps to derive mass-size data on large spatial scales. In G11,
the structure on very small scales is probed in a similar fashion,
using interferometric dust emission maps obtained using the
Submillimeter Array (SMA; Pillai et al., in prep.). Unfortunately, the
SMA map does not cover all of G11, and the derived mass-size data are
likely to be biased (i.e., the most dense core is not covered by our
data). We do not use our existing SMA data for G10, since the data is
misleading in this context (our G10 map only covers a minor dust
emission peak). In the SCUBA and SMA maps, structure larger than a
certain spatial scale is removed during data reduction. Setting other
uncertainties aside, the true intensities and column densities will
therefore be larger than derived here. The adopted dust temperatures
are inspired by \citet{pillai2006:irdc-ammonia} and
\citet{pillai2007:initial-conditions}.

\subsubsection{Mass Estimates}
Table \ref{tab:included-data} lists (as part of the data column) the
dust temperatures adopted for each cloud. The temperatures for
individual fragments may deviate from these mean values by several
Kelvin. This introduces an associated uncertainty in mass estimates
from dust emission. An uncertainty of order 25\% is probably
reasonable for extreme cases.

For Taurus, Figure \ref{fig:cloud-sample} reveals a jump of order of a
factor 2 in the mass-size relation from dust extinction to dust
emission data, with the dust emission data implying the higher
masses. This suggests a problem in the relative calibration of
tracers, or a problem with the generation of the extinction
map\footnote{The other extinction maps used here---created by Lada,
  Alves, Lombardi, and collaborators---use a spatially constant
  resolution. Also, the calculation of the extinction for a given star
  differs in details.} Here, we use the Taurus extinction map with
caution in our analysis.

B59 of the Pipe Nebula map appears to suffer from even larger mass
biases; for exactly the same region, when compared to the
\citet{roman-zuniga2009:b59-extinction} results, the lower-resolution
\citet{lombardi2006:pipe} map implies masses lower by a factor
$\sim 3$. Given the extreme column densities in the target region
(B59), this result is not entirely surprising. We do not expect such
biases in the other regions (where areas of high column density are
anyway probed by dust emission). Still, the B59 results suggest to use
extinction maps with extreme caution.

\section{Mass-Size Relations: A Cloud Sample}
At this point, all mass-size data has been collected and
processed. In the following sections, we highlight three particular
trends seen in this sample data.\label{sec:sample-analysis}

In principle, the mass-size data for a given cloud might significantly
depend on the viewing direction. Since we can only observe the cloud
projections as seen from earth, our data might be biased because of
our specific viewing direction. Here, we make the assumption that this
is not the case. This notion is supported by our observational
findings: all nearby clouds not forming massive stars have (in a broad
sense) similar properties. Also, mass-size differences between clouds
correlate with projection-independent cloud properties (e.g., for given
radius, clouds forming massive stars are more massive). This suggests
that projection effects do not significantly bias the mass-size laws
derived here.

\subsection{A limiting Mass-Size Relation for Massive Star
  Formation?\label{sec:mass-size-trends}}
\begin{figure}
\begin{tabular}{lc}
\includegraphics[width=0.88\linewidth,bb=15 53 360 387,clip]{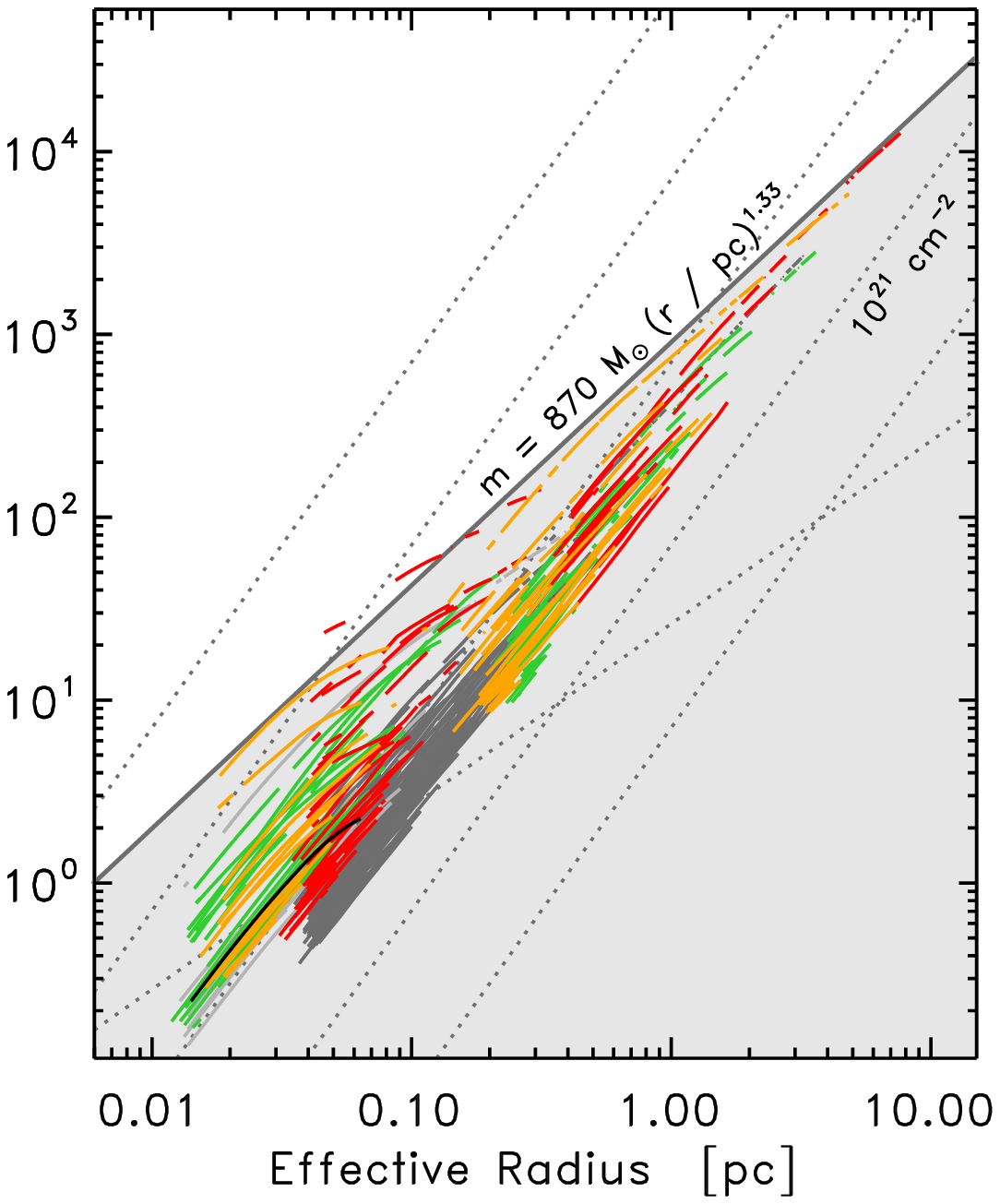} &
\begin{sideways}
\textsf{\hspace{1.8cm}\large{}(a) no high-mass stars}
\end{sideways}\\
\includegraphics[width=0.88\linewidth,bb=15 11 360 387,clip]{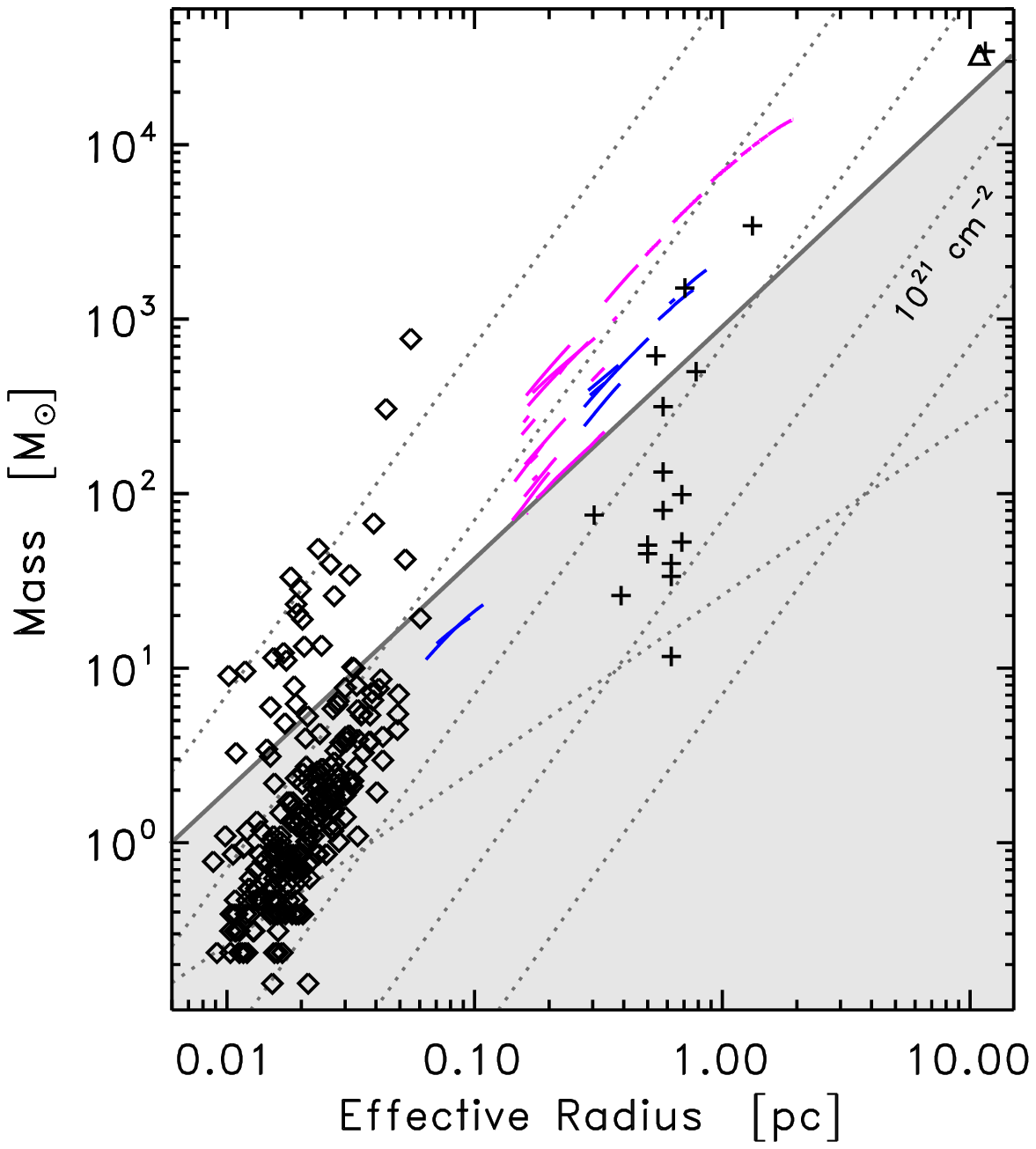} &
\begin{sideways}
\textsf{\hspace{2.5cm}\large{}(b) high-mass stars forming}
\end{sideways}
\end{tabular}
\caption{Joint mass-size plot for all clouds, separated into clouds
  with (\emph{bottom}) and without (\emph{top}) active formation of
  high mass stars. See Fig.\ \ref{fig:cloud-sample} for color
  coding. The \emph{shading} indicates the approximate mass limit for
  fragments in clouds without high mass star formation (\emph{top
    panel}; Eq.\ \ref{eq:mass-reference_limit-low-mass}). In our
  sample, all clouds forming high mass stars contain fragments that
  exceed this limit (\emph{bottom}). See Sec.\
  \ref{sec:mass-size-trends} for
  details.\label{fig:mass-size-comparison}}
\end{figure}

As a first characterization of the cloud sample, we look at the
maximum radius-dependent mass of cloud fragments. We begin with clouds
not forming massive stars. In the
$0.01 \lesssim r / {\rm pc} \lesssim 10$ radius range, essentially all
fragments in such clouds have a mass smaller than some limiting law,
\begin{equation}
m(r) = 870 \, M_{\sun} \, (r / {\rm pc})^{1.33} \, .
\label{eq:mass-reference_limit-low-mass}
\end{equation}
Figure \ref{fig:mass-size-comparison} gives an illustration. This
limit thus gives the typical mass range for structure in clouds like
Taurus, Perseus, Ophiuchus, and the Pipe Nebula.

In detail, this relation excludes a bright fragment in NGC1333, which
we remove from our analysis as a possibly unphysical outlier (its
estimated mass exceeds those of other fragments of similar size by a
factor 2, possibly because of neglected protostellar heating). The
limiting law is derived by searching for the smallest intercept for
which $m_0 \, (r / {\rm pc})^b$ does still provide an upper limit to
the data. Practically, this is done by varying $b$ until
$\max[m(r) / (r / {\rm pc})^b]$ is minimized.\medskip

Interestingly, our sample clouds with massive star formation exceed the
limiting relation (Eq.\ \ref{eq:mass-reference_limit-low-mass}). This is
shown in Fig.\ \ref{fig:mass-size-comparison}(b). For the Orion~A
cloud and G10, we derive an excess of up to a factor 10 in the
$0.01 \lesssim r / {\rm pc} \lesssim 2$ radius range. This suggests
that these clouds have a structure significantly different from what
is found for clouds not containing such stars (Fig.\
\ref{fig:mass-size-comparison}[a]). In this light, Eq.\
(\ref{eq:mass-reference_limit-low-mass}) may approximate a limit for
massive star formation: it could be that only clouds also containing
fragments that exceed Eq.\ (\ref{eq:mass-reference_limit-low-mass})
are able to form massive stars. Larger samples of clouds forming
massive stars must be screened to prove this point. Note, however,
that most fragments in Orion~A do fulfill Eq.\
(\ref{eq:mass-reference_limit-low-mass}). This cloud does therefore
also contain objects that have masses and sizes not distinguishable
from those found for clouds not forming massive stars.

As a mass-size limit for clouds not forming massive stars, Eq.\
(\ref{eq:mass-reference_limit-low-mass}) is probably uncertain to just
a few 10\%. If we, e.g., not use Ophiuchus data in the derivation of
Eq.\ (\ref{eq:mass-reference_limit-low-mass}), then Ophiuchus would
exceed the resulting mass-size limit by 30\%. If we do the same with
Perseus data, the excess is 15\%. It is plausible to expect similar
changes for other regions not forming massive stars. Observational
uncertainties are most likely of a similar order, if one adopts the
mass measurement techniques used here (i.e., dust emission and
extinction). Then, several uncertainties (e.g., dust opacities) are
simply removed by calibration to the same standard. In an absolute
sense (i.e., when considering the true masses), Eq.\
(\ref{eq:mass-reference_limit-low-mass}) is as accurate as the mass
conversion standards used here (e.g., relation of dust emission and
mass). These are probably uncertain to less than a factor 2.

We stress that, excluding observational uncertainties, Eq.\
(\ref{eq:mass-reference_limit-low-mass}) provides a strict upper mass
limit to the solar neighborhood sample \emph{as we have defined it
  here}. Clouds violating Eq.\
(\ref{eq:mass-reference_limit-low-mass}) are not similar to the clouds
in the solar neighborhood sample discussed here. This statement is
sufficient for many purposes.

\subsection{Local Clouds obey similar Mass-Size Relations at large
  Scales}
Inspection of Figs.\ \ref{fig:cloud-sample} and
\ref{fig:mass-size-comparison} suggests that all solar neighborhood
clouds have similar masses at given size for radii $\gtrsim 1 ~ \rm pc$.
Specifically, the most massive fragments at given radius are
essentially all within $\pm 30\%$ of
\begin{equation}
m(r) = 400 \, M_{\sun} \, (r / {\rm pc})^{1.7}
\label{eq:mass-reference_large-radii}
\end{equation}
in the $1 \le r / {\rm pc} \le 4$ radius range, when considering
Taurus, Ophiuchus, Perseus, and the Pipe Nebula. Also our preliminary
Orion~A data is, at $11 ~ \rm pc$ radius, within 40\% of this
law---with the caveat that we cannot examine whether Orion~A follows
Eq.\ (\ref{eq:mass-reference_large-radii}) down to $\sim 1 ~ \rm pc$
radius. It thus appears that---with Orion~A as a possible
exception---all local clouds are similar in their large-scale mass
structure.

This mass-size relation is very similar to the one originally derived
by \citet{larson1981:linewidth_size},
$m(r) = 460 \, M_{\sun} \, (r / {\rm pc})^{1.9}$ (see Eq.\
\ref{eq:mass-size-larson} above). Our study thus confirms his
result---but only for spatial scales $\gtrsim 1 ~ \rm pc$. As we show
throughout this paper (e.g., Eq.\
[\ref{eq:mass-reference_limit-low-mass}] and Figs.\
\ref{fig:global-slopes_sf} and \ref{fig:slope-comparison}), no single
mass-size relation describes all structural aspects of our
observational data.

Some cloud fragments $\gtrsim 1 ~ \rm pc$ are, however, much more
massive than solar neighborhood clouds of similar size. G10, for
example, violates Eq.\ (\ref{eq:mass-reference_large-radii}) by about
an order of magnitude. The similarity between local clouds implied by
Eq.\ (\ref{eq:mass-reference_large-radii}) might thus only hold for
the solar neighborhood.

\subsection{Clusters vs.\ Isolated Star
  Formation\label{sec:clustered-isolated}}
It is obvious that the formation of a cluster requires a larger mass
reservoir than necessary to form a single isolated star. Thus, one
might naively expect that regions containing clusters are more massive
than those devoid of stellar groups. If true, one would thus expect
that, within a given cluster-forming cloud, the regions containing
clusters are more massive than cluster-less regions of similar
size. This hypothesis is tested in Sec.\
\ref{sec:clusters-dominate-host}. Also, one would expect that
cluster-forming cloud fragments are more massive than all
similar-sized fragments in clouds not containing clusters. Section
\ref{sec:clusters-dominate-isolated} investigates this issue.

In our sample, the Pipe Nebula and Taurus serve as examples of regions
dominated by isolated star formation. Actually, except for B59, the
Pipe Nebula does hardly form stars at all
\citep{forbrich2009:pipe-yso}. Perseus and Ophiuchus serve as examples
for cluster-forming regions. They do contain clusters much more
significant than any stellar aggregate found in Taurus and the Pipe
Nebula\footnote{Clusters in nearby clouds were recently surveyed by
  \citet{gutermuth2009:cluster-survey}. Based on Spitzer data,
  Ophiuchus and Perseus are found to contain clusters with
  $\gtrsim 130$ members. The same study gives $\sim 40$ members for
  L1495 in Taurus. This latter value is in line with previous studies
  and sets an upper limit to the size of stellar groups in Taurus
  (Table 4 of \citealt{kenyon2008:sf-handbook}). In the Pipe Nebula,
  the star formation activity is dominated by the B59 region
  \citep{forbrich2009:pipe-yso}. This group has $\sim 20$ members
  \citep{brooke2007:b59}. Taurus and the Pipe Nebula do thus not
  contain clusters as significant as the ones in Perseus and
  Ophiuchus.}. Orion~A is another example of a cluster-forming cloud.

\subsubsection{Cluster-forming Fragments dominate their Host
  Cloud\label{sec:clusters-dominate-host}}
We examine whether cluster-forming cloud fragments dominate the mass
reservoir of their host cloud at all radii. This is executed in Fig.\
\ref{fig:cloud-sample}. Here, we only consider cluster-forming clouds
with high quality data, i.e.\ Perseus and Ophiuchus. Inspection of the
column density maps reveals that the (highlighted) most massive
small-scale features in Perseus and Ophiuchus are located in the
NGC1333 and L1688 clusters, respectively. At small spatial scales, the
most massive fragments are thus indeed located in cluster-forming
regions. Note, though, that cluster-forming regions do also contain
fragments of lower mass. This is illustrated in the rightmost panel of
Fig.\ 6 in part I: Bolocam-detected fragments in NGC1333 cover a wide
range in mass.

In Ophiuchus, when examining larger spatial scales, fragments
containing L1688 continue to be the most massive ones. This is not
exactly true for NGC1333 in Perseus, though. In this cloud, when
considering extinction maps, other fragments (drawn in red) are---by a
small margin---the most massive ones at given radius. However, closer
inspection (not presented here in detail) reveals that these fragments
all contain the cluster IC348. In summary, cluster-forming fragments
do thus indeed constitute the most massive cloud features at given
radius. Some cluster-bearing fragment, which constitutes the most
massive cloud fragment at some radius, might however at a different
radius be less massive than some other cluster-forming fragment. In
this context, note that IC348 is probably much older than NGC1333
(e.g., \citealt{gutermuth2009:cluster-survey}). Dense star-forming
cores, which manifest as small objects of large mass, are thus
actually not expected to remain in the IC348 region.

\begin{figure}
\includegraphics[width=\linewidth,bb=20 10 361 386,clip]{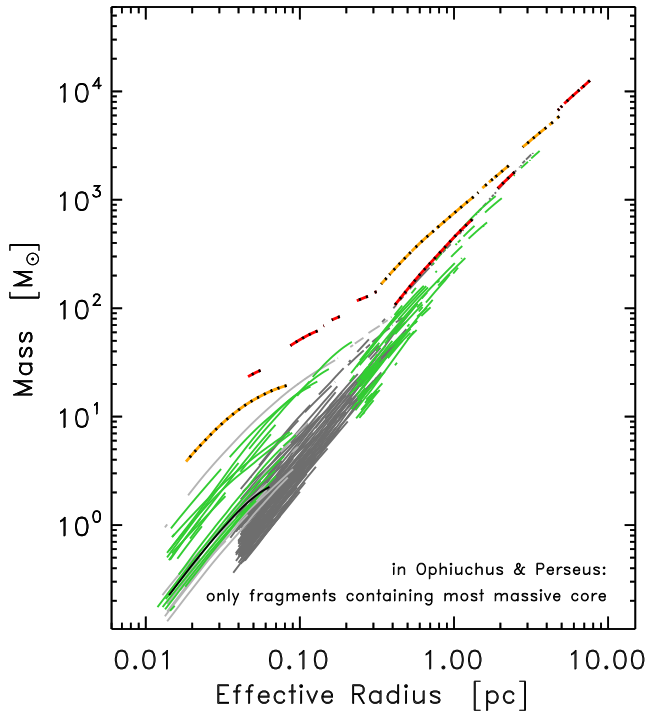}
\caption{Perseus and Ophiuchus cloud regions containing the most
  massive cloud fragment found for small radii (highlighted by
  \emph{dotted black lines}) in comparison to structure in Taurus and
  the Pipe Nebula. See Fig.\ \ref{fig:cloud-sample} for color
  coding. Section \ref{sec:clustered-isolated} describes that
  cluster-forming fragments exceed, at given size, those without
  clusters in mass.\label{fig:mostmassive}}
\end{figure}

\subsubsection{Cluster-forming Fragments exceed Fragments without
  Clusters in Mass\label{sec:clusters-dominate-isolated}}
Figure \ref{fig:mostmassive} compares cluster-forming cloud fragments
to clouds devoid of significant clusters. For Taurus and the Pipe
Nebula, the mass-size data is presented as done before (e.g., Fig.\
\ref{fig:cloud-sample}). We stress again that the Taurus MAMBO data
for small spatial scales does not cover the entire cloud. \emph{At
  $r \lesssim 0.2 ~ \rm pc$, the data only characterize the conditions
  in regions devoid of significant stellar groups.} For Perseus and
Ophiuchus, we choose a different plotting scheme. Here, we only plot
the data for fragments containing the most massive fragment found at
small radii (NGC1333 in Perseus, L1688 in Ophiuchus).

We find that, at given radius, cluster-forming Ophiuchus cloud
fragments (i.e., towards L1688) are significantly more massive than
those in Taurus and Pipe. The naive expectation (i.e., that
cluster-forming fragments are more massive) is thus confirmed. For
Perseus (i.e., towards NGC1333), however, the result is more
nuanced. At $r \lesssim 0.3 ~ \rm pc$, cluster-forming fragments are
significantly more massive than any structure in Taurus or the Pipe
Nebula. For $0.4 \lesssim r / {\rm pc} \lesssim 2$, however, the
extinction-derived mass towards NGC1333 is similar to the maximum
extinction-based masses for Taurus and the Pipe Nebula. If true, this
would suggest that some cluster-forming regions at
$r \gtrsim 0.4 ~ \rm pc$ have a structure similar to clouds only
forming isolated stars (or no stars at all).

It may be, though, that the Perseus extinction-based masses are biased
towards lower values. The Perseus map has relatively poor physical
resolution ($5 \arcmin$ at $260 ~ \rm pc$ distance), and intrinsic
colors of stars in NGC1333 can bias extinction measurements. These
problems are amplified by high column densities towards
NGC1333. Section \ref{sec:data-analysis} demonstrates these
problems for B59. Thus we speculate that, at given size, the
region containing NGC1333 is indeed more massive than any structure in
Taurus and the Pipe Nebula. This remains to be proven, though.

The fragments in the Orion Nebula Cluster, as well as Orion~A as an
entire cloud are more massive than any feature in Taurus and the Pipe
Nebula (Fig.\ \ref{fig:mass-size-comparison}). These data are thus
consistent with the aforementioned trend.

In summary, fragments containing clusters appear to be more massive
than all structure in clouds devoid of clusters. Larger samples, and
better data, are needed to ultimately establish this trend.

\subsection{Global Slopes of cluster-forming
  Regions\label{sec:global-slopes_sample}}
As we just have seen in Sec.\ \ref{sec:clusters-dominate-host},
cluster-forming regions dominate the mass reservoir of their host
cloud at any given spatial scale. This suggests a tight correlation
between the properties of the cluster-forming fragments and the
large-scale cloud structure. This can, for example, be characterized
by ``global'' slopes of connection lines between mass-size
measurements of the most massive fragments at large and small scale
(Eq.\ \ref{eq:slope-global}). Since we lack comprehensive data on all
spatial scales, we are unfortunately presently unable to derive global
slopes for G10 and G11. To do so, one must also carefully characterize
flux losses due to spatial filtering, as they occur in bolometer and
interferometer maps. This is beyond the scope of the present paper.

\begin{figure}
\includegraphics[height=\linewidth,angle=-90]{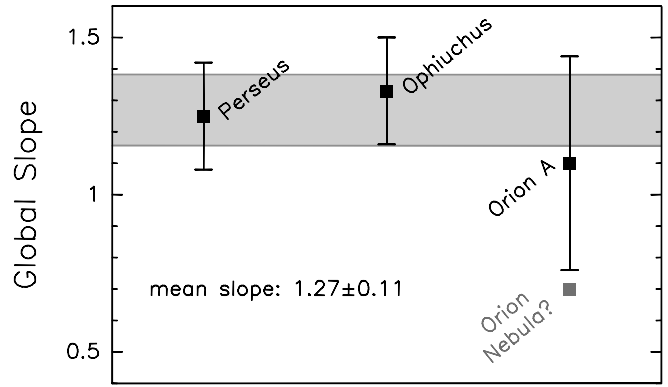}
\caption{Global slopes for cluster-forming clouds. The slopes refer to
  the trends indicated in Fig.\ \ref{fig:cloud-sample}; the
  calculation of uncertainties is discussed in the text. The
  uncertainty-weighted mean slope of the sample is
  $1.27 \pm 0.11$. All clouds are consistent with having this slope,
  but the individual slope uncertainties are significant. If we
  include (highly uncertain) Orion Nebula data, slopes as low as
  $\sim 0.7$ are found. See Sec.\ \ref{sec:global-slopes_sample} for
  details.\label{fig:global-slopes_sf}}
\end{figure}

Specifically, the cluster-forming clouds are probed both at
$\sim 0.05 ~ \rm pc$ and $\sim 5.0 ~ \rm pc$ radius (e.g., Fig.\
\ref{fig:cloud-sample}). We calculate their global slopes by
connecting the most massive fragments detected within 10\% of these
reference radii (30\% radius deviation for the sparsely sampled Orion
data). For Orion, where only very large scales are probed reliably, we
use a mass measurement at $11 ~ \rm pc$ radius for slope
calculations. Also, in Orion, the $\gg 100 \, M_{\sun}$ fragments at
$\sim 0.05 ~ \rm pc$ radius (right in the center of the Orion
Nebula) are rejected as outliers. This rejection
of objects is a regrettable move, since cloud fragments with extreme
properties could be the actual sites where the most massive stars are
born. However, we feel that a more detailed and careful analysis than
possible here, including a detailed consideration of the temperature
structure, is warranted. We hope to do this in one of our future
studies. The resulting slopes range from 1.10 to 1.33, as shown in
Fig.\ \ref{fig:global-slopes_sf}. Note that a much smaller slope (as
small as $\sim 0.7$) would hold for Orion~A, if we do not reject the
outliers. The uncertainties indicated in Fig.\
\ref{fig:global-slopes_sf} hold for an uncertainty of a factor 2 in
the mass ratio (factor 4 for the less reliable Orion data). This error
budget presents the worst expected scenario; the true errors are
probably smaller. In Ophiuchus, we have sufficient resolution to
also measure masses at $\sim 0.02 ~ \rm pc$ radius. If we use such
higher resolution data for this cloud, the slope decreases to 1.30.

Note that we experimented with different approaches to define global
slopes to characterize the core-cloud structure. We then decided to
use the current method using masses only at $0.05 ~ \rm pc$ and
$5.0 ~ \rm pc$ because we find no significant differences compared
with the other methods. For example, following the
construction of Eq.\ (\ref{eq:mass-reference_limit-low-mass}), we
measure for each cloud the slope, $b$, that minimizes
$\max[m(r) / (r / {\rm pc})^b]$. This value is then derived by
utilizing data at all spatial scales. Slopes derived in this fashion
only differ by $\pm 0.04$ from those shown in Fig.\
\ref{fig:global-slopes_sf}, though. In the end, Eq.\
(\ref{eq:slope-global}) appears to provide the simplest approach that
can also be repeated by other researchers.

The uncertainty-weighted mean global slope for the three regions is
$1.27 \pm 0.11$. As shown in Fig.\ \ref{fig:global-slopes_sf}, our
data are consistent with the hypothesis that all cluster-forming
clouds have the same global slope. This suggests that mass-size
relations $m \propto r^{1.27 \pm 0.11}$ provide a crude tool to use
mass measurements at large spatial scale ($\sim 5 ~ \rm pc$) to gauge
the star formation conditions of the most massive embedded regions
($\sim 0.05 ~ \rm pc$). Unfortunately, the extreme SCUBA cores in
Orion~A, with masses $\gg 100 \, M_{\sun}$, suggest that much smaller
slopes may hold in some clouds. The data are incompatible with larger
slopes, though. Thus, the uncertainties shown in Fig.\
\ref{fig:global-slopes_sf} suggest that the slopes are not larger than
1.5.

\subsection{Tangential Slopes: A Transition to Dense
  Cores?\label{sec:slopes-tangential-sample}}
\begin{figure}
\includegraphics[width=\linewidth,bb=19 8 367 278,clip]{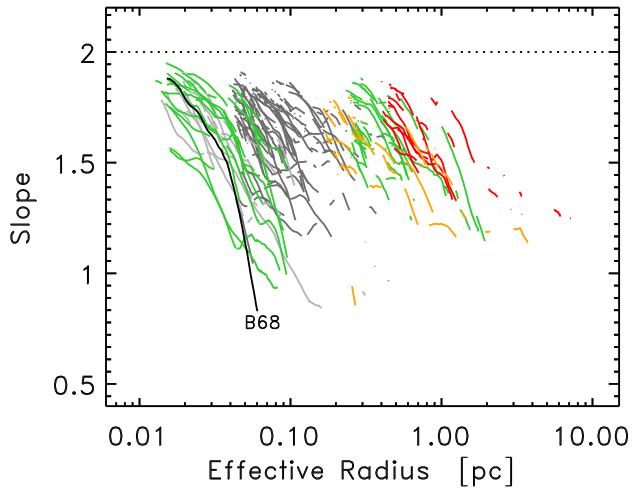}
\caption{Tangential slopes for the sample clouds. See Fig.\
  \ref{fig:cloud-sample} for mark-up and coloring. The \emph{dotted
    line} indicates the upper slope limit inherent to our measurement
  technique (Sec.\ \ref{sec:slope-definitions}). Because of spatial
  filtering, slopes from Taurus MAMBO data may be biased for $r
  \gtrsim 0.1 ~ \rm pc$; such data are removed here. Data from other
  bolometers are not drawn because of the same problem. The
  \emph{black line} gives the slope for B68, a Bonnor-Ebert-like dense
  core; similar slope trends are expected for other cores resembling
  Bonnor-Ebert spheres. See Sec.\ \ref{sec:slopes-tangential-sample}
  for details.\label{fig:slope-comparison}}
\end{figure}

Figure \ref{fig:slope-comparison} presents tangential slopes for the
entire sample. Because of too strong spatial filtering, we exclude
maps taken with interferometers, SCUBA, and Bolocam from this
calculation. For MAMBO, Taurus fragments larger $r = 0.1 ~ \rm pc$
might be affected by filtering
\citep{kauffmann2008:mambo-spitzer}. Their slopes are thus not drawn.

A first trend noted in Fig.\ \ref{fig:slope-comparison} is that the
infinitesimal slopes change with radius. Also, they have no obvious
relation to the global slopes shown in Fig.\
\ref{fig:global-slopes_sf}. Both slope definitions are thus
complementary.

At given spatial scale $\gtrsim 0.5 ~ \rm pc$ (in this domain, we only
have data for Perseus, Taurus, and Ophiuchus), all clouds have
comparable infinitesimal slopes. Those for Ophiuchus are a bit lower
than for the other regions, indicating that the observed clouds differ
in their structure at $r \gtrsim 0.5 ~ \rm pc$. In all clouds the
slopes decrease with increasing radius. As we detail in Sec.\
\ref{sec:density-pressure}, this behavior resembles the mass-size
trends for model clouds with finite extent (i.e., density vanishes
outside a finite radius).\medskip

In Taurus and the B59 and B68 regions of Ophiuchus, we derive very
small infinitesimal slopes for $r \lesssim 0.1 ~ \rm pc$. (These
regions are the only ones well probed at such small scales.) Near
$r = 0.1 ~ \rm pc$, some fragments are observed to have slopes as low
as ${\rm d} \, \ln(m) / {\rm d} \, \ln(r) \sim 1$. This result is not
entirely unexpected. Spheres supported by isothermal pressure, for
example, obey ${\rm d} \, \ln(m) / {\rm d} \, \ln(r) = 1$ at
intermediate radii (Sec.\ \ref{sec:density-pressure}). Such
(Bonnor-Ebert) spheres are commonly believed to constitute good
idealizations of dense core structure. In particular, B68 is today
quoted as the textbook model of a Bonnor-Ebert sphere
\citep{alves2001:b68}. As expected, B68 shows a slope approaching 1
for large radii. We caution that the spatial filtering of bolometer
maps, as well as the removal of extended features in extinction maps,
might bias the slope measurements towards lower values. However, at
least for cores with simple geometries, like B68, slopes $\sim 1$ are
expected for larger radii. This is required for Bonnor-Ebert-like
density profiles. If such slopes were not observed, the paradigm of
Bonnor-Ebert-like dense cores (e.g.,
\citealt{bergin2007:dense-core-review}) would not be consistent with
our data.

Note that the Taurus and Perseus infinitesimal slopes at
$r \sim 0.5 ~ \rm pc$ are much larger than the slopes expected for the
embedded dense cores of simple geometry ($\sim 1$ for every core at
some radius). If this dichotomy is real, it has interesting
implications. Basically, the slope must decrease from
$\gtrsim 1.5$ at $r \gtrsim 0.5 ~ \rm pc$ to
${\rm d} \, \ln(m) / {\rm d} \, \ln(r) \sim 1$ for radii
$\lesssim 0.1 ~ \rm pc$. In other terms, there appears to be a
transition in slope from the diffuse cloud structure to
self-gravitating dense cores, when considering decreasing radii. Again
we must caution that these trends might partially reflect artifacts in
the data. As demonstrated before (Sec.\ \ref{sec:data-analysis}),
the mass in extinction maps is sometimes increasingly underestimated
towards small scales. This biases slopes towards lower values. It is
not likely, though, that all features in our extinction maps are
massively biased. In particular for Perseus, such would
be inconsistent with comparisons between CO and extinction maps
(\citealt{goodman2008:column-density}; also see part I). Still, it is
discomforting that the slope transition does apparently exactly occur
in the spatial domain that is not well probed by any tracer.

Interestingly, a cloud-to-core slope transition might not exist in
Ophiuchus: note that slopes $\sim 1$ are observed for radii
$\gtrsim 0.3 \rm pc$, so that no slope transition between cloud and
core might be necessary. Specifically, at any given radius, the most
massive fragment containing L1688 follows tightly the line connecting
the most massive fragments at $0.05 ~ \rm pc$ and $5 ~ \rm pc$
radius. If this is true, a clean division between the L1688 dense
cores and the surrounding diffuse cloud structure might not exist.

\section{Interpretation of Mass-Size Data}
\label{sec:interpretation}

\subsection{Larson's 3$^{rd}$ Law and other previous
  Work\label{sec:comparison-prev-work}}
\citet{larson1981:linewidth_size} carried out one of the first studies
of mass-size relations of molecular clouds (in his Fig.\ 5) and
obtained $m(r) \propto r^{1.9}$. (Actually, he studied size-density
relations. Since he assumed spherical cloud geometries, these can be
turned into mass-size laws.) Today, this ``third Larson law'' is
usually quoted as $m(r) \propto r^2$ and considered one of the
fundamental properties of cloud structure (e.g.,
\citealt{mckee2007:review}).

We find a nuanced relation between our data and the detailed result of 
\citet{larson1981:linewidth_size},
$$ m(r) = 460 \, M_{\sun} \, (r / {\rm pc})^{1.9} $$
(our Eq.\ \ref{eq:mass-size-larson}). In the
$1 \le r / {\rm pc} \le 4$ radius range, his and our results are
compatible for several solar neighborhood clouds, as shown by Eq.\
(\ref{eq:mass-reference_large-radii}). However, some clouds, like G10,
can deviate by about an order of magnitude in mass from this
relation. Further, some trends within individual clouds are not
described well by one single mass-size law. Slope and intercept of
Eq.\ (\ref{eq:mass-reference_limit-low-mass}), and the slopes shown in
Figs.\ \ref{fig:global-slopes_sf} and \ref{fig:slope-comparison}, may
serve as examples for deviant mass-size relations. The
\citet{larson1981:linewidth_size} mass-size law thus fails to describe
a lot of the cloud substructure visible in state-of-the-art cloud
maps. Deviations from this relation have also been noted before (e.g.,
\citealt{mckee2007:review} for references). Also, note that
\citet{larson1981:linewidth_size} never intended to match all cloud
substructure; as seen in his Fig.\ 5, the mass-size relation only
provides an order-of-magnitude fit to his data. In this sense, the
mass-size laws derived here do not really supersede Larson's results;
molecular clouds are very complex structures, and every one has to be
considered individually. However, for certain mass-size trends, Eq.\
(\ref{eq:mass-reference_limit-low-mass}) and the global slopes in
Fig.\ \ref{fig:global-slopes_sf} do provide a more detailed
description than offered by Eq.\ (\ref{eq:mass-size-larson}).\medskip

Beyond these quantitative differences, Larson's technique differs
fundamentally from ours. He plotted all cloud and dense core data
available to him, and attempted to fit this ensemble data by a common
law, independent of whether the cores and clouds were residing in the
same area of the sky. Here, however, we usually handle data separately
for each cloud, to e.g.\ calculate the global and tangential slopes
shown in Figs.\ \ref{fig:global-slopes_sf} and
\ref{fig:slope-comparison}. Only Eqs.\
(\ref{eq:mass-reference_large-radii},
\ref{eq:mass-reference_limit-low-mass}) stem from an analysis of data
combined for several clouds.

Ensemble-based mass-size relations of the kind considered by Larson
have been studied by many other groups; here we can only present
excerpts from such work. \citet{lada2008:pipe-nature}, for example,
derive $b = 2.6$ for dense cores in the Pipe Nebula extinction map
used here. For a sample of Infrared Dark Clouds (IRDCs),
\citet{ragan2009:irdc-extinction} derive even steeper slopes,
$b \approx 3$, much in excess of any slope derived here. Their
approach is very different from ours, though; they extract dense
cores, plot their masses and sizes, and fit the data with a single
line. We, instead, measure a core's mass at various contours, and
derive tangential slopes by only considering data for a single dense
core. Their ensemble slopes thus serve a different purpose than our
global and tangential slopes. In this respect, it may be helpful to
consider the Type 1--4 linewidth-size relations of
\citeauthor{goodman1998:coherent_cores}
(\citeyear{goodman1998:coherent_cores}; replace line width with mass
in their discussion): there are many different ways to define slopes,
and they will characterize different properties. The large-scale
structure of molecular clouds, as originally addressed by
\citet{larson1981:linewidth_size}, is in principle discussed by
\citet{solomon1987:scaling-relations} and
\citet{heyer2009:scaling-relations}. Their data has, however, not been
exploited to derive mass-size relations.

Note that \citet{stuewe1990:structural-analysis} uses a
technique similar to ours to probe the spatial domain considered
here. As discussed in Sec.\ \ref{sec:density-pressure}, his work is
consistent with ours.

\subsection{Mass-Size Relations and Star
  Formation\label{sec:slope-star-formation_discussion}}
It is thought since long that clouds need to achieve a high column
density in order to produce dense cores and
stars. \citet{johnstone2004:extinction_threshold_ophiuchus}, in
particular, introduced the concept of extinction (or column density)
thresholds for dense core formation (also see:
\citealt{onishi1998:c180-taurus}, \citealt{hatchell2005:perseus},
\citealt{enoch2007:cloud_comparison}). Similarly,
\citet{lombardi2006:pipe} and \citet{lada2009:california-cloud} argue
that a low fraction of mass at high column density yields low star
formation activity (also see
\citealt{kainulainen2009:column-density-pdf}). This is in line with
the Kennicutt-Schmidt law between star formation rate and mass surface
density, $\Sigma_{\rm SFR} \propto \Sigma_{\rm gas}^p$ (e.g.,
\citealt{kennicutt1998:ks-law}). Since $p \sim 1$ for star-forming
clouds \citep{evans2009:c2d-summary}, this relation predicts an
increase of star formation activity with increasing column density.

The analysis in Sec.\ \ref{sec:sample-analysis} shows that these laws
do also manifest in our data. Sections
\ref{sec:mass-size-trends} and \ref{sec:clustered-isolated} suggest
that the ability to form clusters and massive stars increases with
increasing mass, when considering a given radius. Since
$\langle N_{\rm H_2} \rangle \propto m / r^2$, one could also say that
cloud fragments do appear to only form massive stars and clusters when
they have a high mean column density---just as suggested by the
aforementioned laws.\medskip

Mass-size studies do, however, also provide information not available
from the aforementioned plain column density studies. First, note that
``extinction threshold'' studies (e.g.,
\citealt{johnstone2004:extinction_threshold_ophiuchus}) do only
consider a single spatial scale, i.e.\ the beam used to construct the
extinction map. This is a major difference to mass-size studies, where
many spatial scales are considered. Second, observe that studies of
column density distributions (PDFs; e.g., \citealt{lombardi2006:pipe})
\emph{do} consider all scales of a map, but do not register which part
of the signal shown on the histogram originates is which part of the
cloud. (Depending on the analysis, this is not necessarily a problem.)
Mass-size studies, in contrast, treat individual cloud fragments
separately.

Mass-size data sets give a new twist to discussions of extinction
thresholds. To see this, consider the solar neighborhood clouds
examined here. These are all similar at large spatial scale (Eq.\
\ref{eq:mass-reference_large-radii}). Still, at smaller scale, they
differ significantly in mass and star formation activity (Sec.\
\ref{sec:clustered-isolated}). The processes determining the star
formation activity must thus operate on spatial scales smaller than
the entire cloud. In this sense, \emph{the star formation activity
  depends} (at least in our sample clouds) \emph{on a cloud's ability
  to create, from a given mass reservoir, a small number of fragments
  that dominate the mass reservoir and concentrate it into a small
  volume}. The presence of high column densities are then a
consequence of the cloud structure, but not the governing reason for
the formation of dense cores and stars.

\subsection{Slopes and Intercepts: Constraints on Density and
  Physical Cloud Models\label{sec:density-pressure}}
The observed power-law-like mass-size relations,
$$ m(r) = m_0 \, (r / {\rm pc})^b \, , $$
are characterized by slopes, $b$, and intercepts, $m_0$. As we show
here, slopes and intercepts can be used to gauge densities and their
gradients. At the same time, it is possible to constrain the absolute
level of pressure, and the nature of its origin.

Consider, for example, an infinite equilibrium sphere with power-law
density profile that is supported by isothermal pressure from gas at
temperature $T_{\rm g}$. Then,
\begin{equation}
m(r) = 2.6 \, M_{\sun} \,
\left( \frac{T_{\rm g}}{10 ~ \rm K} \right) \,
\left( \frac{r}{0.1 ~ \rm pc} \right)
\label{eq:mass-size-sis}
\end{equation}
(\citealt{kauffmann2008:mambo-spitzer}, Eq.\ [13], in their case
$\epsilon \to \pi/2$). Thus, if this model holds, the intercept
encodes information on the gas temperature supporting the
cloud. Conversely, the slope predicted by the model, $b = 1$, can be
used to validate the model; if observations yield $b \neq 1$, then the
model does not apply.

It has to be kept in mind that we consider mass-size laws derived from
two-dimensional maps. These are related to, but not identical with,
mass-size laws obtained from three-dimensional density maps. This is
illustrated by the experiments conducted by \citet{shetty2009:ppp-ppv}
who use the fragment identification technique also used by us. Their
analysis is based on three-dimensional numerical simulations of
turbulent clouds. As part of their experiments, they derive mass-size
slopes from their data. For their particular set of simulations, the
power-law slopes derived in this fashion are similar to the number of
dimensions used for mass measurements (i.e., 3 when based on density,
and 2 when based on column density). This underlines that the number
of dimensions considered has to be kept in mind.

\subsubsection{Density Laws}
The above discussion can be extended to include many more models of
cloud structure. We do this in two Appendices. The results of
this analysis are summarized here.

First, let us examine the connection between mass-size laws and cloud
density profiles. Consider a sphere with about constant density for
radii smaller than some flattening radius, $s_0$, a power-law drop at
intermediate radii, $n(s) \propto s^{-k}$ (where $s$ is the distance
from the center), and vanishing density beyond some outer truncation
radius, $R$. Such profiles provide a good match to observations of
dense cores \citep{tafalla2002:depletion}. They are a good
approximation to the structure of isothermal equilibrium spheres
\citep{dapp2009:density-profile}. As we show in Appendix
\ref{sec-app:density-spheres}, for apertures of radius $r$, this
yields mass-size relations of the form
\begin{equation}
m(r) \propto n_{\rm c} \, r^{3 - k}
\quad {\rm for} \quad
s_0 \ll r \ll R \, ,
\end{equation}
where $n_{\rm c}$ is the density for $s = 0$, and mass-size
slopes
\begin{equation}
{\rm d} \, \ln(m) / {\rm d} \, \ln(r) = 3 - k
\quad {\rm for} \quad
s_0 \ll r \ll R \, .
\end{equation}
Both relations apply only if $k < 3$. As the equations show, the
intercept contains information on the central density, and the
mass-size slope depends on the slope of the density law, $k$. Both
does, of course, only hold at intermediate radii, $s_0 \ll r \ll R$.
For reference, we note that the column density obeys
$N \propto r^{1 -   k}$.

A generalized version of power-law spheres are tri-axial
ellipsoids. Appendix \ref{sec-app:density-triaxial} considers the case
in which $n(s) \propto (s / s_0)^{-k}$ along any main axis, but with
$s_0$ depending on the direction chosen (Eq.\
\ref{eq-app:density-law-ellipsoidal}). Detailed analysis shows that
such ellipsoids follow the same mass-size relations as spheres, when
$r = (A / \pi)^{1/2}$. Thus, the laws listed above apply.

In a next step, one may wish to consider models of cylindrical clouds
of length $\ell$. Here, we adopt density drops $n(s) \propto s^{-k}$
perpendicular to the cylinder axis for intermediate values of $s$.
At intermediate radii $s_0 \ll r \ll R$, such clouds obey
\begin{equation}
m(r) \propto n_{\rm c} \, r^{4 - 2 k} \quad \Rightarrow \quad
{\rm d} \, \ln(m) / {\rm d} \, \ln(r) = 4 - 2 k
\label{eq:slope-cylinder_perp}
\end{equation}
if their major axis is perpendicular to the line of sight, and
\begin{equation}
m(r) \propto n_{\rm c} \, r^{2 - k} \quad \Rightarrow \quad
{\rm d} \, \ln(m) / {\rm d} \, \ln(r) = 2 - k \, ,
\label{eq:slope-cylinder_par}
\end{equation}
if the axes are aligned. (Intermediate angles are not considered
here.) Meaningful relations are only obtained for $k < 2$. In both
relations, the radius is defined as $r = (A / \pi)^{1/2}$.

Thus, just as one may naively expect, the mass-size slope gauges the
slope of the density profile. Further, the intercepts of mass-size
relations constrain the absolute density of cloud fragments. There is,
however, one less obvious fact that calls for attention: the exact
relations between mass-size slopes, intercepts, and density law
depends on the cloud model and viewing angle. It is therefore not
possible to derive the true density profile without further
information on the cloud geometry. Such information may, e.g., be
derived by studying the elongation of cloud fragments. Also, the
above power-law relations do only apply at intermediate radii,
$s_0 \ll r \ll R$. This domain might not exist in actual observed
clouds. Then, the central density plateau and the finite size have to
be considered. These give
\begin{equation}
m(r) \left\{
\begin{array}{llll}
\propto n_{\rm c} \, r^2 & {\rm for} & r \ll s_0 &{\rm and}\\
\approx M & {\rm for} & r \gtrsim R & .\\
\end{array}
\right.
\end{equation}

\subsubsection{Polytropic Equilibria}
The density slopes themselves depend on the processes shaping the
model cloud. As a first example, here we consider static equilibrium
models in which pressure gradients are in balance with
self-gravity. We assume a polytrophic equation of state,
$P \propto n^{\gamma_P}$, in which pressure and density are
related by the polytrophic exponent, $\gamma_P$.

In Appendix \ref{sec-app:polytropes} we show that
\begin{equation}
k = \frac{2}{2 - \gamma_P}
\label{eq:slope-gammap}
\end{equation}
for polytropic equilibrium spheres (if $\gamma_P < 4/3$) and
cylinders (if $\gamma_P < 1$). The density and mass-size slopes
are, thus, related to the polytrophic exponent. Isothermal pressure,
for which $\gamma_P = 1$, implies $k = 2$ in spheres, for example.
Then, ${\rm d} \, \ln(m) / {\rm d} \, \ln(r) = 1$ in spherical model clouds;
laws too complex to be considered here apply to cylinders. As seen in
Figs.\ \ref{fig:global-slopes_sf} and \ref{fig:slope-comparison}, such
a model can explain some, but not most slope measurements.

Polytropic exponents $\gamma_P = 1/2$ are sometimes suggested to
describe ``turbulent'' pressure within clouds, as e.g.\ arising from
Alfv\'en waves \citep{mckee1995:alfven_waves}. In this case,
$k = 4/3$, and so ${\rm d} \, \ln(m) / {\rm d} \, \ln(r)$ assumes values of
$5/3 \approx 1.67$ (spheres and ellipsoids), $4/3 \approx 1.33$
(perpendicularly viewed cylinder), and $2/3 \approx 0.67$ (end-on
cylinder) for the different models. Among these, spheres, ellipsoids,
and cylinders viewed from the side provide an acceptable
match to the observed mass-size slopes $b > 1$. Cylinders viewed along
their major axis yield too shallow mass-size laws (and such a
viewing direction is highly unlikely).

For a given physical model, the intercept can be used to gauge a
cloud's stability against collapse, respectively suggest the level of
supporting pressure. Here, we limit ourselves to the isothermal case, i.e.\
$\gamma_P = 1$. Stability considerations (e.g., of Bonnor-Ebert-type;
\citealt{ebert1955:be-spheres}, \citealt{bonnor1956:be-spheres}) imply
\begin{equation}
M \le M_{\rm cr} \approx 2 \, \frac{\sigma^2(v) \, R}{G}
\label{eq:stability}
\end{equation}
for the total mass, where $\sigma(v)$ is the characteristic
one-dimensional velocity dispersion (Eqs.\
\ref{eq-app:mass-limit-spheres} and
\ref{eq-app:mass-limit-cylinders}). For spheres, $R$ is the radius,
while $R \to \ell$ in cylinders. If $\sigma(v)$ is known (e.g., for
thermal pressure), $M > M_{\rm cr}$ implies collapse of the object
considered. Conversely, depending on the situation, $\sigma(v)$ can be
inferred by requiring that $M = M_{\rm cr}$. Required values of
$\sigma(v)$ significantly in excess of the thermal velocity dispersion
of the mean free particle might, e.g., suggest the presence of
significant non-thermal pressure. If we only require that pressure
balances gravity, and drop the constraint that the object is to be
stable against perturbations, the above law yields Eq.\
(\ref{eq:mass-size-sis}).

As particular example, consider B68. It is thought that this dense
core has a structure very similar to a Bonnor-Ebert sphere
\citep{alves2001:b68}. Thus, one would expect the mass and size of B68
to obey Eq.\ (\ref{eq:mass-size-sis}), when considering large enough
radii. This is indeed the case, as seen in Fig.\ \ref{fig:cloud-sample}.

\subsection{Synoptic and physical Density Slopes}
For intuitive communication, it may be helpful to report synoptic
density slopes,
\begin{equation}
\left[ - \frac{{\rm d} \, \ln(n)}{{\rm d} \, \ln(s)} \right]_{\rm syn} =
3 - \frac{{\rm d} \, \ln(m)}{{\rm d} \, \ln(r)} \, ,
\end{equation}
i.e.\ the density slope a sphere of the same mass-size slope would
have when observed at intermediate radii. The synoptic slopes give a
good first idea of the true density slopes. First, recall that the
model mass-size laws do not sensitively depend on the assumption of
exact spheres; the same relation holds for ellipsoids. Also, in the
observed range $1 \lesssim {\rm d} \, \ln(m) / {\rm d} \, \ln(r) < 2$,
spheres (or ellipsoids) and perpendicularly viewed cylinders (the
end-on view is statistically insignificant) imply similar slopes; for
these geometries, the synoptic slopes exceed the true ones by less
than 0.5, assuming intermediate radii. Thus, we derive
$$ \left[ - {\rm d} \, \ln(n) / {\rm d} \, \ln(s) \right]_{\rm syn} =
1 ~ {\rm to} ~ 2 $$
for mass-size slopes $1 \lesssim {\rm d} \, \ln(m) / {\rm d} \, \ln(r) < 2$.

A limited comparison of these density-size slopes with previous
results is possible. \citet{tafalla2002:depletion}, e.g., study the
dust emission of five starless dense cores, and derive density-size
slopes of $2.0 ~ {\rm to} ~ 2.5$ for four of them. This is a typical
result for cloud fragments of $\lesssim 0.1 ~ \rm pc$ size
\citep{bergin2007:dense-core-review}, a spatial domain not too well
covered by our data.

Concerning the analysis method and spatial range considered, the
extinction study by \citet{stuewe1990:structural-analysis} might
provide the best match to our work. Based on star counts, he derives 
$- {\rm d} \, \ln(n) / {\rm d} \, \ln(s) > 1.0 \pm 0.4$ on scales of up to
$\sim 1 ~ \rm pc$. This is consistent with our results, also given the
differences in map construction (he uses optical star counts).

\section{Summary \& Outlook}
This work studies the internal structure of molecular clouds by
breaking individual cloud complexes up into several nested
fragments. For these, we derive masses and sizes in order to study
their density structure.\label{sec:summary}

Analysis of a limited sample of solar neighborhood cloud complexes
(Taurus, Ophiuchus, Perseus, Pipe Nebula, Orion~A), as well as more
distant clouds (G10 \& G11), yields first some basic constraints on
mass-size cloud structure. These are as follows.
\begin{enumerate}
\item On large spatial scales, $\ge 1 ~ \rm pc$, the most massive
  fragments in solar neighborhood clouds---with the possible exception
  of Orion~A---obey
  $$ m(r) = 400 \, M_{\sun} \, (r / {\rm pc})^{1.7} $$
  (Eq.\ \ref{eq:mass-reference_large-radii}) with deviations
  $< 40 \%$. This relation resembles the original
  mass-size law derived by \citet{larson1981:linewidth_size},
  $m(r) = 460 \, M_{\sun} \, (r / {\rm pc})^{1.9}$. The more distant
  clouds in the sample, however, deviate from this relation by up to
  an order of magnitude in mass.
\item No single mass-size law can be used to describe all fragments in
  all clouds. In particular, ``Larson's 3$^{\rm rd}$ Law'' of constant
  column density, $m(r) \propto r^2$, provides a bad global
  description; today's data are too complex to warrant the use of such
  relations. To give examples, power-law slopes vary with radius
  within a given cloud (Fig.\ \ref{fig:slope-comparison}), and clouds
  can differ massively in mass at given radius (Fig.\
  \ref{fig:mass-size-comparison}).

  In practice, different definitions of mass-size laws are used by
  different researchers. Also, different definitions may serve
  different purposes. This must be taken into account when comparing
  different mass-size laws.
\end{enumerate}
Most importantly, the mass-size data can be used to learn about the
formation of stars in molecular clouds. We derive the following
constraints.
\begin{enumerate}
\setcounter{enumi}{2}
\item Sample clouds not forming massive
  stars ($\gtrsim 10 \, M_{\sun}$) adhere to a limiting mass size
  relation,
  $$ m(r) \le 870 \, M_{\sun} \, (r / {\rm pc})^{1.33} $$
  (Eq.\ \ref{eq:mass-reference_limit-low-mass}),
  while our sample clouds forming such stars violate this law (Fig.\
  \ref{fig:mass-size-comparison}). This suggests that the above
  relation describes the typical mass-size range of molecular clouds
  not forming high-mass stars. Also, the observations advocate that
  this boundary constitutes a mass limit for massive star
  formation. However, such conclusions are based on a small sample and
  are thus preliminary.
\item Across all clouds studied here, cloud fragments forming clusters
  are more massive than fragments not doing so (Figs.\
  \ref{fig:cloud-sample} and \ref{fig:mostmassive}; Sec.\
  \ref{sec:clusters-dominate-host}). At given size, cluster-forming
  fragments dominate the mass reservoir of their host cloud.
\item The mass-size trend of cluster-forming fragments can e.g.\ be
  captured by global mass-size slopes (i.e., from $0.05 ~ \rm pc$ to
  $5.0 ~ \rm pc$ radius; Sec.\ \ref{sec:global-slopes_sample}). Our
  cluster-forming sample clouds are consistent with a common slope
  $\sim 1.27$. The uncertainties are, unfortunately, significant for a
  given cloud; slopes may well differ between clouds. In the case of
  Orion~A, e.g., the slope might be as low as $\sim 0.7$. However, in
  any event slopes smaller 1.5 do hold.
\end{enumerate}
Theoretical discussions show that mass-size laws of the form
$m(r) = m_0 \, (r / {\rm pc})^b$ can be related to physical cloud
models characterized by power-law density gradients,
$n(s) \propto s^{-k}$, or polytropic equations of state,
$P \propto n^{\gamma_P}$ (Sec.\ \ref{sec:density-pressure}). Provided
certain idealizations apply, $b$, $k$, and $\gamma_P$ are directly
related to another. This analysis suggests the definition of a
synoptic density slope,
$ \left[ - {\rm d} \, \ln(n) / {\rm d} \, \ln(s) \right]_{\rm syn} =
3 - {\rm d} \, \ln(m) / {\rm d} \, \ln(r) $
(i.e., assuming the fragment considered was a sphere). This slope
provides a first rough estimate of the true density law. Our data
gives synoptic density slopes in the range $1 ~ {\rm to} ~ 2$.

\acknowledgements{We are indebted to a careful and thorough anonymous
  referee, who saved us from making an embarrassing mistake. This
  project would not have been possible without help from Erik
  Rosolowsky. His dendrogram analysis code
  \citep{rosolowsky2008:dendrograms} was instrumental for our
  analysis. \citet{carey2000:irdc-submillimeter}, Alves et al.\
  (priv. comm.), \citet{enoch2006:perseus}, \citet{lombardi2006:pipe},
  \citet{ridge2006:complete-phase_i},
  \citet{roman-zuniga2009:b59-extinction}, and Thompson et al.\ (in
  prep.) contributed maps to the present study. We are grateful for
  their help. This work was in part made possible through Harvard
  Interfaculty Initiative funding to the Harvard Initiative in
  Innovative Computing (IIC). It is based upon work supported by the
  National Science Foundation under Grant No.\ AST-0908159.}

\bibliographystyle{apj}
\bibliography{mendeley/library}

\appendix

\section{Model Mass-Size Relations}
Here we consider density profiles of the form
\begin{equation}
n(s) \propto s^{-k}
\end{equation}
in the domain $s_0 \ll s \ll R$. In spheres, $s$ is the distance from
the density peak, while it is the distance from the main axis in
cylinders. Densities become about constant for $s \lesssim s_0$, and
they vanish for $s > R$.\label{sec-app:mass-size-models}

\subsection{Spheres at Intermediate Radii\label{sec-app:density-spheres}}
Integration along the line of sight gives the column density at offset
$u$,
\begin{equation}
N(u) = 2 \,
\int_0^{[R^2 - u^2]^{1/2}} n([u^2 + w^2]^{1/2}) \, {\rm d} \, w \, ,
\label{eq-app:cd-integration}
\end{equation}
where the integration stops at the boundary radius. Evaluation yields
\begin{equation}
N(u) \propto u^{1 - k} \quad (R \to \infty)
\end{equation}
if $k > 1$, a relation that can be used for offsets $s_0 \ll u \ll
R$. Integration of the column density over a circular peak-centered
aperture of maximum offset $u$ gives the mass,
\begin{equation}
m(u) = 2 \pi \mathcal{C} \int_0^u N(w) \, w \, {\rm d} \, w \, ,
\label{eq-app:coldens-to-mass}
\end{equation}
where $\mathcal{C} = \Sigma / N_{\rm H_2} = \mu_{\rm H_2} m_{\rm H}$
is the conversion factor to mass surface density. Evaluation in the
limit $R \to \infty$ gives
\begin{equation}
m(u) \propto u^{3 - k} \quad \Rightarrow \quad
m(r) \propto r^{3 - k}
\end{equation}
for $s_0 \ll r \ll R$. We use $u = r$ in the transition from offsets
to aperture sizes since, in spheres,
the offset from the center, $u$, is equal to the radius defined via
the aperture area, $r = (A / \pi)^{1/2}$.

The meaningful range of this relation is limited to $k < 3$; for
$k \ge 3$, the mass would decrease with size, which is not physical.
In essence, spheres with $k \ge 3$ have a finite mass, even for
$R \to \infty$, so that power-law mass-size relations cannot hold in
any radius domain.

\subsection{Homogeneous Spheres\label{sec-app:density-hom_spheres}}
Homogeneous spheres, in which the density is spatially constant,
constitute a special case of the aforementioned spherical
models. Consider an offset $u$ from the sphere's center, where the
density drops to zero beyond the sphere's outer radius, $R$. Then, the
column density is the product of the density, $n_0$, and the length of
the line of sight,
\begin{equation}
N(u) = 2 n_0 (R^2 - u^2)^{1/2} \, .
\end{equation}
Integration following Eq.\ (\ref{eq-app:coldens-to-mass}) yields (with
the substitution $u \to r$)
\begin{equation}
m(r) = \frac{4}{3} \pi \varrho_0 \,
\left[ R^3 - (R^2 - r^2)^{3/2} \right] \, ,
\end{equation}
in which $\varrho_0 = \mathcal{C} n_o$ is the mass density
corresponding to $n_0$. For $r \to R$ we derive
$m = 4/3 \pi \varrho_0 R^3$, just as expected for a homogeneous sphere
truncated at radius $R$. The slope obeys
\begin{equation}
\frac{{\rm d} \, \ln(m)}{{\rm d} \, \ln(r)} =
3 \, \frac{
  (R^2 - r^2)^{1/2} \, r^2
}{
  R^3 - (R^2 - r^2)^{3/2}
}
\end{equation}
and does monotonously decrease from
${\rm d} \, \ln(m) / {\rm d} \, \ln(r) = 2$ at $r = 0$ to
${\rm d} \, \ln(m) / {\rm d} \, \ln(r) = 0$ at $r = R$.

\subsection{Triaxial Ellipsoids at Intermediate
  Radii\label{sec-app:density-triaxial}}
Another variation of spherical power-law density profiles are triaxial
density distributions with ellipsoidal iso-density surfaces. These can
be derived from spherical power-laws by substituting
$s / s_0 \to [(x / x_0)^2 + (y / y_0)^2 + (z / z_0)^2]^{1/2}$  (where
a subscript `0' indicates reference properties for normalization). The
coordinates $x$, $y$, and $z$ give the projection of a given position
on the three axes of the ellipsoidal density distribution. Then, the
density law reads
\begin{equation}
n = n_0 \,
\left[
\left( \frac{x}{x_0} \right)^2 +
\left( \frac{y}{y_0} \right)^2 +
\left( \frac{z}{z_0} \right)^2
\right]^{-k/2} \, .
\label{eq-app:density-law-ellipsoidal}
\end{equation}
The chosen coordinates form an orthogonal coordinate system. Any
position along a straight line of sight, $w$, is therefore linearly
related to the coordinates chosen for the density distribution:
\begin{equation}
x = m_x + n_x \, w \, ; \quad
y = m_y + n_y \, w \, ; \quad
z = m_z + n_z \, w \, .
\label{eq-app:coordinate-transforms-ellipsoid}
\end{equation}
For convenience, here we choose that $w = 0$ at the position of
highest density along the line of sight. Substitution of Eqs.\
(\ref{eq-app:coordinate-transforms-ellipsoid}) into Eq.\
(\ref{eq-app:density-law-ellipsoidal}) yields
\begin{equation}
n = n_0
\left[
\frac{m_x^2}{x_0^2} + \frac{m_y^2}{y_0^2} + \frac{m_z^2}{z_0^2}  +
2 \, \left(
\frac{m_x n_x}{x_0} + \frac{m_y n_y}{y_0} + \frac{m_z n_z}{z_0}
\right) \, w +
\left(
\frac{n_x^2}{x_0^2} + \frac{n_y^2}{y_0^2} + \frac{n_z^2}{z_0^2}
\right) \, w^2
\right]^{-k/2} \, ,
\label{eq-app:density-law-ellipsoidal-substituted}
\end{equation}
i.e., some terms that do not contain $w$, some that include $w$
linearly, and a few that contain $w$ to its square. Analysis shows
that the sum of the terms with a linear dependence on $w$ must be zero.
To see this, consider the requirement that the density reaches its
maximum where $w = 0$. For this to happen, the derivative of the sum
within the square brackets of Eq.\
\ref{eq-app:density-law-ellipsoidal-substituted} with respect to $w$ must
vanish at $w = 0$. This is only the case if all terms linear in $w$
cancel out. Equation (\ref{eq-app:density-law-ellipsoidal-substituted})
can thus be written
\begin{equation}
n = n_0 \,
\left( \frac{u^2 + w^2}{r_0} \right)^{-k / 2} \, ,
\label{eq-app:density-law-ellipsoidal-simplified}
\end{equation}
if one chooses
$u^2 = (m_x^2/x_0^2 + m_y^2/y_0^2 + m_z^2/z_0^2) \cdot
(n_x^2/x_0^2 + n_y^2/y_0^2 + n_z^2/z_0^2)^{-1}$
and
$r_0^2 = (n_x^2/x_0^2 + n_y^2/y_0^2 + n_z^2/z_0^2)^{-1}$.
Formally, Eq.\ (\ref{eq-app:density-law-ellipsoidal-simplified}) is
identical to the spherical density law substituted in Eq.\
(\ref{eq-app:cd-integration}). Therefore, we can use the calulations
in Appendix \ref{sec-app:density-spheres} to conclude that, again,
\begin{equation}
N(u) \propto u^{1 - k} \quad (R \to \infty)
\end{equation}
if $k > 1$. Thus, the spatial column density distribution does only
depend on $u$.

A detailed look at the above analysis does actually show that we did
not use the exact nature of the density law to this point. Instead, we
exploited that the density depends on
$(x / x_0)^2 + (y / y_0)^2 + (z / z_0)^2$. Thus, we can make the
conclusion that \emph{all} density distributions with ellipsoidal
iso-density surfaces yield column density distributions only depending
on $u$.

Thus, $u$ can be interpreted as an offset, just as in the spherical
situation. Inspection of the equation defining $u$ reveals that the
three-dimensional offsets from the distribution's center, $m_i$, form
a triaxial ellipsoid too, when $u$ is kept constant. Projection of the
$u = const.$ surface into a plane corresponds to a cut through this
ellipsoid. Since cuts though ellipsoids yield ellipses, the
$u = const.$ curve in a plane is an ellipse too. In this case,
spatial integration of the column density can be conveniently executed
in polar coordinates. Let $v$ and $w$ be the minor and major axis of
the projected ellipse. For a given viewing perspective, $v / w$ is
constant for all contours of constant column density. Then, the mass
follows from
\begin{equation}
m(u) = 2 \pi \, (v/w) \, \mathcal{C}
\int_0^u N(w) \, w \, {\rm d} \, w \, ,
\end{equation}
where we use that the ellipse's area increases with $w$ as
 $2 \pi \, (v/w) \, w \, {\rm d} \, w$. This integral becomes
\begin{equation}
m(u) \propto u^{3 - k} \quad \Rightarrow \quad
m(r) \propto r^{3 - k}
\end{equation}
for $s_0 \ll r \ll R$. In this, we use that $u$ is proportional to the
effective radius of the ellipsoid formed by $N = const.$ (i.e.,
$u = const.$) contours. The latter holds since, for a given ellipsoid,
$v/w$ is a constant, and its area is $\pi v w = \pi (v/w) w^2$. Thus, the
effective radius becomes $r = (v/w)^{1/2} w \propto w$.

In summary, we retrieve the relations already derived for spheres,
including the limits on reasonable values of $k$. As expected, spheres
form a particular case of ellipsoids with $x_0 = y_0 = z_0$.

\subsection{Cylinders at Intermediate Radii: Perpendicular
  View\label{sec-app:density-cylinders_perp}}
In cylinders with the main axis perpendicular to the line of sight,
the integration in Eq.\ (\ref{eq-app:cd-integration}) must be
evaluated with the integration path perpendicular to the cylinder
axis. Then
\begin{equation}
N(u) \propto u^{1 - k} \quad (R \to \infty) \, ,
\end{equation}
as long as $k > 1$. This provides, again, a convenient description for
$s_0 \ll u \ll R$. Since the column density is constant for given
offset, $u$, apertures along a given column density contour have a
size $2 u \ell$, where $\ell$ is the cylinder length. Such an aperture
contains a mass
\begin{equation}
m(u) = 2 \ell \mathcal{C} \int_0^u N(w) \, {\rm d} \, w \, .
\end{equation}
Substitution yields
\begin{equation}
m(u) \propto u^{2 - k} \quad \Rightarrow \quad
m(r) \propto r^{4 - 2 k} \, .
\end{equation}
This relation is applicable if $k < 2$, following the discussion of
finite masses for spheres. In the transition from $u$ to $r$ we use
that $r = (2 \ell u / \pi)^{1/2}$, which gives $u \propto r^2$.

\subsection{Cylinders at Intermediate Radii: Parallel
  View\label{sec-app:density-cylinders_para}}
In cylinders with the main axis aligned with the line of sight,
the integration in Eq.\ (\ref{eq-app:cd-integration}) must be
evaluated along the cylinder axis. In this case,
\begin{equation}
N(u) = n(u) \cdot \ell \, .
\end{equation}
This geometry further implies
\begin{equation}
m(u) = 2 \pi \mathcal{C} \int_0^u N(w) \, w \, {\rm d} \, w \, ,
\end{equation}
which gives
\begin{equation}
m(u) \propto u^{2 - k} \quad \Rightarrow \quad
m(r) \propto r^{2 - k}
\end{equation}
in the usual $s_0 \ll u \ll R$ limit, with the further constrain that
$k < 2$. As in spheres, $u = r$ for contours of constant column
density.

\subsection{Very large and small Radii}
For central flattening of the density profile, the column density
becomes constant for $u \ll s_0$, independent of the model adopted.
Then, the mass in a given aperture is just given by the product of the
mass surface density, which scales with $n_{\rm c} R$, and the
aperture area. For large radii, the mass is equal to the total mass
when considering apertures larger than the object's size. Thus
\begin{equation}
m(r) \left\{
\begin{array}{llll}
\propto n_{\rm c} \, R \, r^2 & {\rm for} & r \ll s_0 & , \quad {\rm and}\\
\approx M & {\rm for} & r \gtrsim R & , \\
\end{array}
\right.
\end{equation}
independent of the adopted model geometry.

\section{Polytropic Equilibrium Cloud Models}
Polytropic cloud models assume an equation of state of the form
\begin{equation}
P = P_0 \, (\varrho / \varrho_0)^{\gamma_P} \, .
\end{equation}
They are in particular used to solve the equation of hydrostatic
equilibrium, $\vec{\nabla} P = - \varrho \, \vec{\nabla} \Phi$, where
the gravitational potential fulfills the Poisson equation,
$\vec{\nabla}^2 \Phi = 4 \pi G \varrho$.\label{sec-app:polytropes}

\subsection{Polytropic Equilibrium Spheres}
Hydrostatic equilibrium spheres supported by polytropic pressure have
recently been summarized by \citet{mckee1999:mpp} and
\citet{curry2000:comp_poly}. In spherical symmetry,
$\vec{\nabla} \to \partial / \partial s$, $\Phi(s) = - G m(s) / s$, and
$m(s) = 4 \pi \int_0^s \varrho(w) \, w \, {\rm d} \, w$. This
differential equation can easily be solved for density profiles
$\varrho(s) \propto s^{-k}$. Solutions are, however, only physical for
$k < 3$; the power-law implies $m(s) \propto s^{3 - k}$, which  only
increases with radius if $k$ is small enough. The hydrostatic equation
implies $s^{- \gamma_P k -1} \propto s^{1 - 2 k}$. This can only be
fulfilled if both exponents to $s$ are identical, which implies
\begin{equation}
k = \frac{2}{2 - \gamma_P} \quad (\gamma_P < 4/3) \, ,
\end{equation}
where the limit on $\gamma_P$ reflects the condition $k < 3$.
Stability against perturbations implies that the mass of the
equilibrium is smaller than some critical mass
\citep{mckee1999:mpp}. For $\gamma_P = 1$, this critical mass is
identical to the Bonnor-Ebert mass,
\begin{equation}
M_{\rm cr} = 2.4 \, \frac{\sigma^2(v) \, R}{G} \, ,
\label{eq-app:mass-limit-spheres}
\end{equation}
where the numerical constant follows the discussion by
\citet{mckee1999:mpp}. Stable clouds have $M < M_{\rm cr}$.

\subsection{Polytropic Equilibrium Cylinders}
\citet{horedt1987:polytropes} presents a summary of hydrostatic
equilibrium cylinders (as well as sheets and spheres) supported by
polytropic pressure. It builds on discussions of the case
$\gamma_P \ge 1$ by \citet{ostriker1964:polytropes}, and analysis of
polytropes with $\gamma_P < 1$ by \citet{viala1974:polytropes}.

The Poisson equation can be tackled using the divergence theorem,
$\int_V \vec{\nabla} \vec{F} \, {\rm d} \, V =
\int_{\partial V} \vec{F} \, {\rm d} \, \vec{A}$. When analyzing this
equation for the volume of an infinitely long cylinder, the ends can
be neglected in the integration over the volume's surface,
$\partial V$. Thus, if the component of $\vec{F}$ perpendicular to the
cylinder surface is constant, and assumes the value $F_{\perp}$ along
it, then $| \int_{\partial V} \vec{F} \, {\rm d} \, \vec{A} | =
2 \pi \ell s F_{\perp}$ for cylinders of length $\ell$ and radius
$s$. To apply this analysis to the Poisson equation, we substitute
$\vec{F} \to \vec{\nabla} \Phi$. In cylindric coordinates, the
component of the potential's gradient that is perpendicular to the
cylinder surface implies the substitution
$F_{\perp} \to \partial \Phi / \partial s$. In essence, combination of
the Poisson equation and the divergence theorem yields
$4 \pi G \, \int_V \varrho \, {\rm d} \, V =
\int_{\partial V} \partial \Phi / \partial s \, {\rm d} \, A$, which
evaluates to $\partial \Phi / \partial s = 2 G m(s) / (s \ell)$, where
$m(s)$ is the mass within the radius $s$. For $\ell \to \infty$, we
further have $\vec{\nabla} \to \partial / \partial s$, and so the
hydrostatic equation becomes
$\partial P / \partial s = - \varrho \, \partial \Phi / \partial s$.

A simple solution is provided by $\varrho(s) \propto s^{-k}$, which
implies $s^{- \gamma_P k -1} \propto s^{1 - 2 k}$, just as found for
the spherical case. Analysis of the integral for the mass implies
$k < 2$, similar to what is found for spheres. Thus,
solutions must fulfill
\begin{equation}
k = \frac{2}{2 - \gamma_P} \quad (\gamma_P < 1) \, ,
\end{equation}
where the limit enforces $k < 2$. A perturbation analysis analog to
the spherical Bonnor-Ebert case gives a critical limiting mass
\begin{equation}
M_{\rm cr} = 2 \, \frac{\sigma^2(v) \, \ell}{G}
\label{eq-app:mass-limit-cylinders}
\end{equation}
for $\gamma_P = 1$, as e.g.\ demonstrated by
\citet{ostriker1964:polytropes}.

\end{document}